\documentclass[aip,cha,preprint,showpacs]{revtex4-1}

\usepackage{graphicx}
\usepackage{dcolumn}
\usepackage{bm}
\usepackage{amsmath}
\usepackage{latexsym}
\usepackage{amsmath}
\usepackage{amssymb,amsmath}

\begin{document}
\title{Synchrony suppression in ensembles of coupled oscillators  via adaptive
vanishing feedback}

\author{Ghazal Montaseri$^{1,2}$, Mohammad Javad Yazdanpanah$^{3}$, Arkady Pikovsky$^{2}$, and
Michael Rosenblum$^{2}$}
\affiliation{$^1$Advanced Control Systems Laboratory, School of Electrical and Computer
Engineering, College of Engineering, University of Tehran, Tehran, Iran\\
$^{2}$Department of Physics and Astronomy, University of Potsdam,
Karl-Liebknecht-Str. 24/25, 14476 Potsdam, Germany\\
$^{3}$Advanced Control Systems Laboratory, Control and Intelligent Processing
Center of Excellence, School of Electrical and Computer Engineering, College of
Engineering, University of Tehran, Tehran, Iran}

\date{\today}

\begin{abstract}
Synchronization and emergence of a collective mode is a general phenomenon, frequently
observed in ensembles of coupled self-sustained oscillators of various natures.
In several circumstances, in particular in cases of neurological pathologies, this state 
of the active medium is undesirable. Destruction of this state by a specially designed 
stimulation is a challenge of high clinical relevance.
Typically, the precise effect of an external action on the ensemble is unknown, since the 
microscopic description of the oscillators and their interactions  are not available. 
We show, that desynchronization in case of a large degree 
of uncertainty about important features of the system is nevertheless possible; 
it can by achieved by virtue of a feedback loop with
an additional adaptation of parameters. The adaptation 
also ensures desynchronization of ensembles with
non-stationary, time-varying parameters. 
We perform the stability analysis of the feedback-controlled  system and demonstrate 
efficient destruction of synchrony for several models, including those 
of spiking and bursting neurons. 
\end{abstract}


\maketitle
\begin{quotation}
Synchronization is a general effect in coupled oscillator systems: due to 
an adjustment of individual rhythms, the objects start to oscillate in tact,
producing a pronounced collective rhythm. This is observed not only in physical and technical systems
(lasers, Josephson junctions, spin-torque oscillators, metronomes) but in many 
biological (e.g. fireflies) and even social (applause) systems. Synchronization
can be influenced by an external forcing of the system. In some situations synchrony is not desirable:
e.g., Parkinson's tremor is attributed to a pathological synchrony in population of neurons in the brain.
One tries to suppress this synchrony by a properly designed external forcing, which is arranged via a proportional 
feedback. In this respect, suppression of synchrony can be treated as a control problem. 
The main difficulty here is that the particular mechanism, of effects of  forcing on
oscillating elements is not known. To overcome this problem, we
suggest here an adaptive scheme which adjusts the parameters of the feedback loop  and, thus, compensates the absence of knowledge about microscopic organization of the ensemble. We demonstrate the feasibility of this approach on several
examples, including also highly non-stationary situations where readjustment of the parameters is needed.
\end{quotation}
\section{Introduction} 

Synchronization is adjustment of rhythms of coupled oscillating objects
due to their weak interaction \cite{Pikovsky_et_al:Book}. In physics, biology,
engineering, and neuroscience, a wide range of synchronization phenomena
in large oscillator populations is known, such as coordinated firing of cardiac 
pacemaker cells, synchronous regimes in
arrays of Josephson junctions and lasers, synchronization in ensembles of
electronic circuits  and in neuronal populations 
\cite{Pikovsky_et_al:Book,Mosekilde_et_al:2002,%
Strogatz:2003,Balanov_et_al:2009}. 

In neuroscience, synchronization of neurons plays an
important role in vital functions like vision, movement control,  and memory \cite{Cumin_Unsworth:2007}.
On the other hand, such diseases as epilepsies, Parkinson's disease, and essential tremor
are believed to be related to a pathologically enhanced synchronization of neurons
\cite{Tass_book:1999,Milton-Jung-03,Buzhaki-Draguhn-04,Batista_et_al:2010,Park_et_al:2011}.  
Many patients suffering from these diseases cannot be cured by known medications
and are therefore treated by Deep Brain Stimulation (DBS) which implies stimulation
of the brain tissue via implanted microelectrodes
\cite{Chkhenkeli-78,Chkhenkeli-03,Benabid_et_al:1991,Kringelbach-07,Bronstein-11}. 
Typically, the stimulation is delivered by a subcutaneously implanted controller.
In its present form, DBS is a permanent open-loop stimulation with high frequency (about 120 Hz) pulses;
in neurological practice the controller's parameters are tuned empirically.
DBS is known to cause serious side-effects like speech problems and involuntary muscle contractions.
Besides, permanent intervention into the brain tissue results in fast discharge of the controller's
batteries, and, consequently, in further minor surgery.
Despite DBS clinical success, its exact mechanism is not  yet completely 
understood~\cite{Hammond_et_al:2008}. Presumably, the effect of high-frequency stimulation is 
not related to desynchronization of neurons, but rather to lesioning of the tissue via 
suppression of the neuronal activity.    

DBS limitations and high clinical relevance have encouraged experimentalists to search for more efficient
stimulation algorithms. So, a feedback-based DBS has been recently tested in a study with primate model 
of Parkinson's disease~\cite{Rosin_et_al:2011} and with rodent model of epilepsy~\cite{Bereny_et_al:2012}.
Besides these empirical studies, there were quite a number of theoretical efforts within the physical and 
engineering community.
The key idea of this activity, initiated by P. Tass~\cite{Tass_book:1999,Tass:2003},  
is that the stimulation should be able to suppress the abnormal 
synchrony  among neurons without putting them to silence, or, in physical terms, 
to desynchronize the synchronized oscillators without quenching them.  
In this approach the neuronal population is typically modeled as a 
network with high connectivity and treated in the mean field approximation.
The mean field of the ensemble is associated with the pathological brain rhythm; 
hence, the goal of the stimulation is to minimize the mean field. 

There are two groups of desynchronizing techniques. The first group is based on the idea 
of phase resetting by precisely timed pulses~\cite{Tass_book:1999,Tass:2003,Lysyansky:2011}, 
while the second group involves the methods from the control theory and relies on continuous 
feedback~\cite{Rosenblum_Pikovsky_a:2004,Rosenblum_Pikovsky_b:2004}. 
The latter is based on the measurement of the mean field which one wants to diminish. 
The feedback may be proportional to the mean field or its delayed value, or be a 
nonlinear function of it~\cite{Rosenblum_Pikovsky_a:2004,Rosenblum_Pikovsky_b:2004,%
Popovych_et_al:2006,Tukhlina_et_al:2007,Pyragas-Popovych-Tass-07,%
Popovych_et_al:2008,Kobayashi_Kori:2009,Luo_et_al:2009,Popovych_Tass:2010,Luo_Xu:2011,%
Franci_et_al:2011,Franci_et_al:2012}. 
Multi-site feedback controllers are considered in 
\cite{Hauptmann_et_al:2007,Omelchenko_et_al:2008,Guo_Rubin:2011}. Feedback control of two interacting 
subpopulations is addressed
in~\cite{Rosenblum-Tukhlina-Pikovsky-Cimponeriu-06,Tukhlina_Rosenblum:2008,Popovych_Tass:2010}.
In the context of the neuroscience application, the crucial feature of the feedback schemes is their 
potential ability to provide \textit{vanishing stimulation control}, i.e. to ensure that the stimulation tends to 
zero (to be exact, to the noise level), as soon as the goal of the control is achieved and the undesired 
synchrony is suppressed~\cite{Rosenblum_Pikovsky_b:2004,Tukhlina_et_al:2007}. 
In the present contribution we extend the results of ~\cite{Tukhlina_et_al:2007}, designing an 
adaptive vanishing stimulation setup.
 
In many applications, in particular in neuroscience, the mechanism of external action
on individual oscillators and their interactions is not fully understood. 
So, when an electric stimulation of the brain tissue is applied via an implanted electrode, many 
factors regarding the impact of the stimulation remain unknown, e.g.,
whether the stimulation affects only the membrane voltage of a cell 
or it may influence the gating variables; next, it is not exactly known how the effect of 
stimulation decreases with the distance from the electrode or how its impact 
changes with time, etc. Thus, the feedback control we want to design shall 
work without good knowledge of the system to be controlled and shall 
exploit only rather general models of emergent collective activity.
It means that the controller shall be able to cope with the uncertainty and, possibly, 
with the time drift of parameters of the system. These considerations motivated us 
to implement an adaptive control strategy and to modify
the stimulation technique proposed in \cite{Tukhlina_et_al:2007} to ensure 
adaptive vanishing stimulation.

%
%

The paper is organized as follows. In Section~\ref{preliminaries} we reformulate the problems 
in terms of the control theory and discuss the required features of the controller. 
In Section~\ref{controller_design} we discuss the design of the feedback controller and in 
Section~\ref{stability_analysis} we analyze its stability. Section~\ref{simulation_results}
presents the examples of synchrony suppression in several models of globally coupled 
oscillators. In Section~\ref{conclusion} we summarize our results. 
Some details of the stability analysis and of numerical simulations are given in Appendices.

\section{Desynchronization as a control problem}
\label{preliminaries}

Consider an ensemble of $N$ coupled self-sustained oscillators. At the
microscopic level, depending on the coupling strength, oscillators may oscillate
incoherently (or asynchronously) or they may show (partially) synchronous
oscillations. At the macroscopic level, i.e. where only the collective motion of
the ensemble is considered,  a transition to synchrony
can be viewed as a Hopf bifurcation, which is described by the normal
form
\begin{equation}\label{uncontrolled_complex_macroscopic_model}
\dot Z = (\xi  + i{\omega _0})Z - {\left| Z \right|^2}Z\;,
\end{equation}
also known as the Stuart-Landau equation.
Here $Z$ and $\omega_0$ are the complex amplitude and frequency 
 of the collective mode (mean field), respectively, and
parameter $\xi>0$ describes the instability 
of the only equilibrium point $Z=0$ of
Eq.~\eqref{uncontrolled_complex_macroscopic_model}.
We emphasize that exact derivation of Eq.~\eqref{uncontrolled_complex_macroscopic_model} 
from the  microscopic dynamics is possible only  in exceptional cases (cf. recent 
papers~\cite{Ott-Antonsen-08,Ott-Antonsen-09}, where 
Eq.~\eqref{uncontrolled_complex_macroscopic_model} has been derived for the Kuramoto 
model of sine-coupled phase oscillators).
For general self-sustained oscillators and general coupling, and especially for  live systems 
where the models, if known, are very approximate, this is not feasible, and 
Eq.~\eqref{uncontrolled_complex_macroscopic_model} remains a 
phenomenological model equation.

At the macroscopic level, onset of synchronous or asynchronous oscillations is
related to the instability or stability of the equilibrium point, respectively. 
 In this framework, the
\textit{desynchronization problem} means designing a stimulation (control input)
$u$, such that the \textit{controlled system}
\begin{equation}\label{controlled_complex_macroscopic_model}
\dot Z = (\xi  + i{\omega _0})Z - {\left| Z \right|^2}Z + {e^{i\beta }}u
\end{equation}
 is stabilized at the origin. 
The phase shift parameter $\beta$ in this model equation reflects the
uncertainty in the impact of the stimulation on oscillators. 
In Ref.~\cite{Rosenblum_Pikovsky_b:2004} some of us have shown that $\beta$
inevitably appears in the normal form equation of the forced globally coupled system;
this parameter  depends on the organization of the global coupling in the ensemble 
and on the properties of individual units (cf. Ref.~\cite{PhysRevLett.97.213902} for a similar
phase shift in an optical feedback).
Therefore, frequently used assumptions that $\beta=0$ or $\beta=\pi$  are not validated by theory and
neglect essential feature of the collective dynamics.
Throughout this paper, we consider $\beta$ as an unknown, possibly time-variant, parameter and 
assume that it may attain all values within  $[0,2\pi )$.

From the control theory point of view, uncertainties modeled by multipliers of
the control input represent the \textit{unknown control directions problem}. Designing a
stabilizing controller for a system with an unknown control direction is more
complicated than for other classes of uncertainties. One intuitive reason is that
unknown control multipliers may change the negative feedback to the positive
one. The solution of this problem is known for several special cases only.  For a scalar system, the solution is obtained using the Lyapunov 
direct method~\cite{Kaloust_Qu:1997} 
or with the help of the iterative learning technique \cite{Xu_Yan:2004}.
For two-dimensional systems, in Ref.\cite{Kaloust_Qu:1995} the stabilizing controller with unknown 
direction is designed if the input $u$
enters one of the equations only.
(If our system \eqref{controlled_complex_macroscopic_model} is re-written 
in coordinates, $X=\mbox{Re}(Z)$, $Y=\mbox{Im}(Z)$, the term 
 ${e^{i\beta }}u$ generally appears in both  equations for $\dot{X}$ and $\dot{Y}$.)   
 This problem is well studied for a class of
nonlinear systems which can be written in the strict feedback form (see e.g.
\cite{Ge_et_al:2008,Liu_Huang:2008,Wen_Ren:2011} and the references therein) or
in the normal form \cite{Bartolini_et_al:2009} 
\footnote{A second order system in the strict feedback form is represented as
$\dot x_1=f_1(x_1)+g_1x_2$, $\dot x_2=f_2(x_1,x_2) + g_2 u$, where $g_{1,2}$ are unknown 
parameters. A $(2+m)^{th}$ order system in the normal form  is represented 
as $\dot x_1=x_2$, $\dot x_2=f(x_1,x_2,y) +g u$, $\dot y=q(x_1,x_2,y)$, where $y\in \mathbb{R}^{m}$ is the vector of additional states (which describe internal dynamics of the feedback loop; see Ref.~\cite{Khalil:2002} for more details)}, 
or if two control inputs $u_{1,2}$ are allowed \cite{Oliveira_et_al:2010}, so that 
factors $\cos\beta$ and $\sin\beta$ can be compensated.
This brief review of the existing techniques to the unknown control direction problem 
shows that they cannot be exploited for our purpose.

In this paper, we  propose a feedback stimulation technique for stabilizing  the 
zero equilibrium point of Eq.~\eqref{controlled_complex_macroscopic_model}. 
The main advantage is that it (i) provides the vanishing control, what reduces the 
intervention into a living tissue and the energy consumption.
Next, the controller (ii) stabilizes the system in the presence of the unknown phase 
shift $\beta$ and (iii) is able to adapt itself to variations of $\beta$.
Having in mind possible neuroscience applications, we also ensure that the designed 
controller has the following properties. It  (iv) performs stabilization using 
the signal which is contaminated by the rhythms produced by
neighboring neuronal populations and the measurement noise. It is able (v)
to  washout  constant component in the measurement. 
Finally, the stimulation (vi) avoids sudden impacts  on the neuronal
ensemble, which may force neurons to behave far from their
natural dynamics, but affects the ensemble gradually and smoothly.

\section{Designing an adaptive stimulation}
\label{controller_design}

To motivate our approach (and for discussion below), we first mention that
the simplest way to linearly stabilize the system
\eqref{controlled_complex_macroscopic_model} at $Z=0$ 
would be to choose $u =  - \gamma {e^{ -i\beta }}Z$, with $\gamma  > \xi $. 
However, the control signal shall be real-valued.  
If we take $u = \mbox{Re} ( - \gamma {e^{ - i\beta }}Z) =  -
\gamma (X\cos \beta + Y\sin \beta)$, the control action will be also stabilizing.
However, since $Y$ is not available and  $\beta$ is unknown, 
this scheme cannot be implemented.
 
\begin{figure*}[ht]
\centerline{\includegraphics[width=\columnwidth]{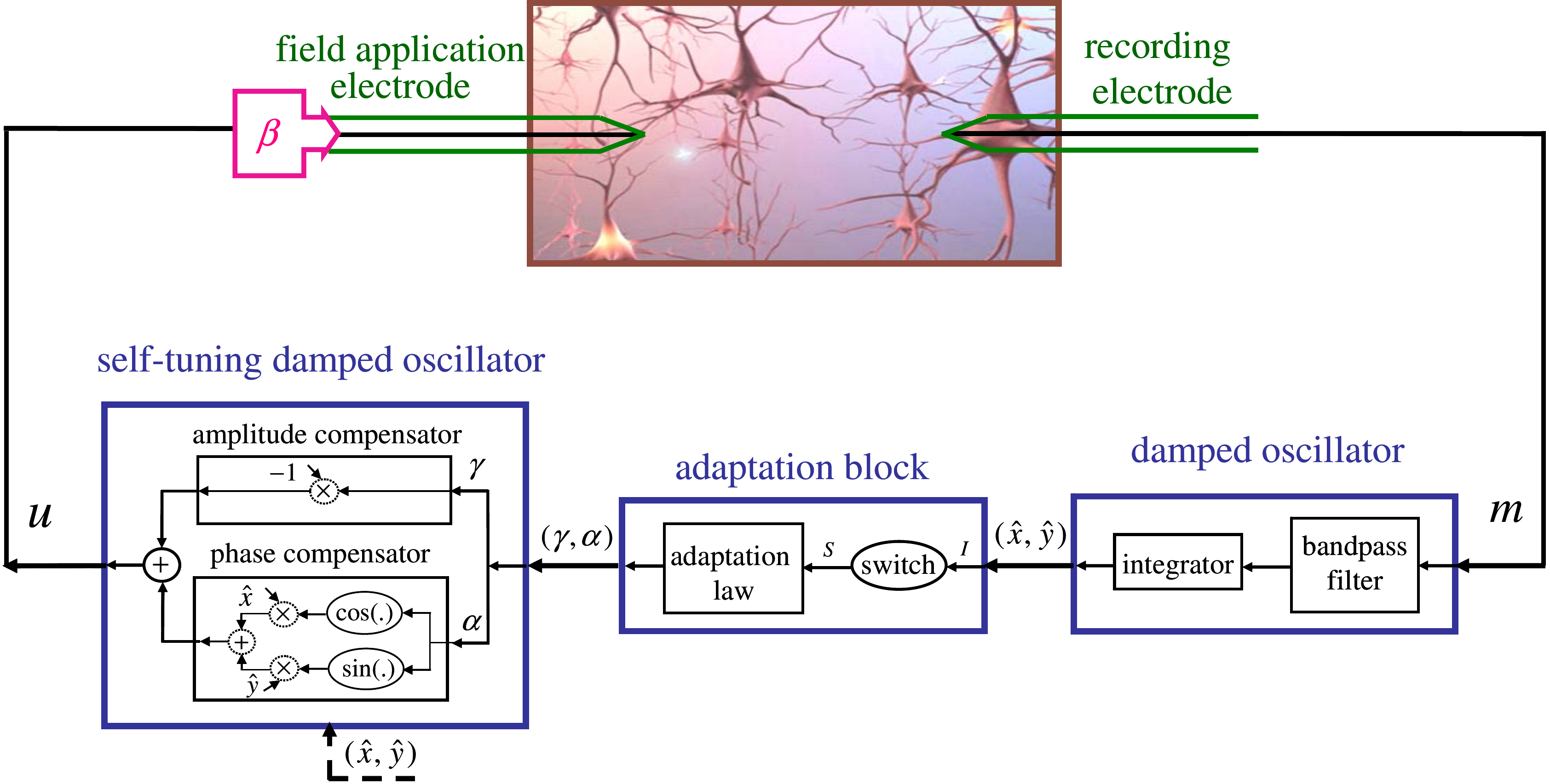}}
\caption{ (Color online) The schematic representation of the controlled neuronal population.
The local filed potential, related to the mean field of the population, is
measured by the recording electrode. The adaptive controller consisting of three
blocks generates the control signal which is fed back to the system via the field
application electrode. The adaptive nature of the stimulation makes it capable
of compensating an \textit{a priori} unknown and slowly varying phase shift parameter $\beta$.}
\label{stimulation_scheme}
\end{figure*}

By constructing the feedback-based stimulation with the mentioned specifications,
we assume that the real-valued measurement $m=Re(Z)+b+n$ is available.
The constant term $b$ reflects the fact that the equilibrium point of the macroscopic 
oscillator is non-zero, what frequently occurs in neuronal models, and $n$ is 
noise. The controller consists of three blocks, as illustrated in Fig.~\ref{stimulation_scheme}, 
where dynamic sub-blocks are shown by squares and static sub-blocks by ellipses.
The first block is described by the equations:
\begin{eqnarray}
  {\dot x}_1& =& {x_2}\;, \label{eq:1bla} \\
  {{\dot x}_2} &=& m - \omega _0^2{x_1} - \delta {x_2}\;, \label{eq:1blb} \\ 
{\dot x}_3 &=& \frac{1}{\mu }({x_2} - {x_3})\;,\label{eq:1blc}
\end{eqnarray}
where Eqs.~(\ref{eq:1bla}) and (\ref{eq:1blb})  constitute a second order bandpass filter which ensures 
accomplishment of the requirements (iv) and (v).
The damping factor $\delta$ determines
the width of the bandpass. Frequency
$\omega_{0}$ is chosen to be close to the basic oscillator frequency of
$Z$, which can be easily measured in the experiment. 
Parameter $\mu$ is chosen so that $\mu {\omega _0} \gg 1$,  then 
the sub-block (\ref{eq:1blc}) operates as an integrator.
Thus, $x_3$  has the same average
as $x_2$ but its phase is shifted by ${\pi }/{2}$.
Finally, we define two \textit{auxiliary oscillating modes} as:
\begin{equation}\label{x_hat_y_hat}
\hat x = \delta {x_2},\,\,\,\,\hat y = \delta\mu\omega_{0}{x_3}.
\end{equation}
 The amplitudes of $\hat x$ and $\hat y$ are close to that of $m$,   
but their phases are shifted by $0$ and ${\pi }/{2}$,
respectively. 

The second block implements an adaptation mechanism. First, we define an observable
$S$ that is proportional to the oscillation amplitude $I = \sqrt {{{\hat x}^2} + {{\hat y}^2}}$, 
with a cutoff at small amplitudes:
\begin{equation}\label{switch}
  S = I\left[ {1 + \tanh \left( {{k_s}(I - {h_s})} \right)}\right] \; .
\end{equation}
Here $S$ is the positive increasing function of $I$; at the cutoff threshold $I\approx h_s$ 
it switches from exponentially small values to $S\approx 2I$, 
the width of the switching
region is governed by $k_s$. The cutoff is required due to the following.  
Although the noise is suppressed in auxiliary oscillating modes $\hat{x}$ and $\hat{y}$, it is
not completely canceled. Therefore the cutoff is
needed to ignore the noise impact below the threshold value $I\approx h_s$.
In a population with a finite size  $N$,
the mean field below the synchronization threshold can be 
treated as a noise with the
root mean square (rms) value proportional to $1/\sqrt N$ \cite{Pikovsky_Ruffo:1999}. So, $I$
is of the order of $\sqrt {2/N}$ which estimates  the  lower bound of $h_s$.

In the next block, the transformed signal $S$ is used to govern the
\textit{adaptive variables} $\alpha$ and $\gamma$ as:
\begin{eqnarray}
  { \dot \alpha}& =& { {k_\alpha
}S,\,\,\,\,\,\,\,\,\,\,\,\,\,\,\,\,\,\,\,\,\,\,\,\,\,\,\,\,\,\,\,\,\,\,\,\,\,\,\,\,\,\ \qquad \alpha (0) = 0}\;, \label{adaptive_variables_alpha} \\
{ \dot \gamma} &=& {{{k_{{\gamma _1}}}S}}
{{\cosh^{-1} \left( {{{{k_{{\gamma _2}}}\gamma }}/
{{{\omega _0}}}} \right)}},\qquad \gamma (0) = 0\;.\label{adaptive_variables_gamma}
\end{eqnarray}
Here $k_{\gamma_{i}},\,i=1,2$ and $k_{\alpha}$ are the positive \textit{adaptation
parameters}. When the measured signal $m$ and  the auxiliary modes 
exhibit large oscillations, $S$ attains some non-zero values. Then, 
 $\alpha$ and $\gamma$ start to grow from zero until the
desired suppression level is achieved and $S$ switches off. Due to the
inevitable  presence of the noise in real applications, if we replace $S$ with
$I$ in Eqs.~(\ref{adaptive_variables_alpha}) and (\ref{adaptive_variables_gamma}), 
the adaptive variables never settle down, because $I$ never vanishes exactly.
Finally, the term $\cosh ({k_{{\gamma _2}}}\gamma /{\omega _0})$ in
Eq.~\eqref{adaptive_variables_gamma} is introduced to suppress the undesired increase of
$\gamma$ which may lead to large control effort and large oscillation amplitude
in the transient time before the stabilization occurs.

Thus, the first and second blocks generate  oscillating modes
$\hat{x},\hat{y}$ and the adaptive variables $\alpha,\gamma$.
The final control signal is in fact the oscillating mode, 
with the phase shifted by $\alpha$
and the amplitude multiplied by $\gamma$:
\begin{equation}\label{control_law}
u =  - \gamma \left( \hat x \cos \alpha + \hat y\sin \alpha  \right).
\end{equation}
This transformation is accomplished in the last block, consisting of a phase shifting 
unit and a multiplier.

Now we recall the discussion in the beginning of
this Section. We see that  the control law Eq.~\eqref{control_law} would
operate properly if  $\hat x\cos\alpha \sim  X\cos\beta$ and 
$\hat y\sin\alpha\sim  Y\sin\beta$.
The similarity between these two controllers gives us an intuitive insight about
the capability  of \eqref{control_law} in stabilizing the zero equilibrium point
of the controlled system  (quantitative analysis will be presented 
in the next Section). It is easy to check that the proposed scheme fulfills all above formulated 
requirements and provides vanishing control, i.e. the maintenance of the desired asynchronous 
state needs no control effort.

One important property that makes the proposed adaptive stimulation distinct
from other proposed techniques is that the stimulation emerges from the origin
(at $t=0$, $\gamma=0$ and thus, $u=0$),  then grows gradually and finally,
settles down again at the origin. During its lifetime, it adapts itself until it
overcomes the natural coupling between the oscillators and thus leads to
desynchronization. Since the stimulation affects the oscillators of the
population gradually and smoothly, it avoids undesired sudden impacts on the
natural behaviors of the stimulated oscillators (requirement (vi)). In the
context of neuroscience, it means that the stimulation intervention into the
living tissue is temporary and smooth (cf. panels showing $u(t)$ in 
Figs.~\ref{van_der_pol_1_beta}-\ref{hindmarsh_burst} 
below).

\section{Stability analysis}
\label{stability_analysis}
The complete set of equations for the closed-loop system reads as:
\begin{eqnarray}\label{closed_loop}
  \dot X &=& \xi X - {\omega _0}Y - X({X^2} + {Y^2})  - \gamma \cos \beta 
  (\hat x\cos \alpha  + \hat y\sin \alpha ) \;,\label{cl_1}\\
  \dot Y &=& {\omega _0}X + \xi Y - Y({X^2} + {Y^2}) - \gamma \sin \beta
  (\hat x\cos \alpha  + \hat y \sin \alpha )\;, \label{cl_2} \\
  {{\dot x}_1} &=& {x_2}\;, \label{cl_3} \\
  {{\dot x}_2} &=& X - \omega _0^2{x_1} - \delta {x_2}\;, \label{cl_4} \\
  {{\dot x}_3} &=& \frac{1}
{\mu }({x_2} - {x_3}) \;,\label{cl_5} \\
 \dot \alpha  &=& {k_\alpha}S \;,\label{cl_6}\\
\dot \gamma  &=& {{{k_{{\gamma _1}}}S}}
{{\cosh^{-1} \left( {{{{k_{{\gamma _2}}}\gamma }}/
{{{\omega _0}}}} \right)}}\;,\label{cl_7}\\
\hat x& =& \delta {x_2},\,\,\hat y = \delta \mu {\omega
_0}{x_3}\nonumber \;,\\  
I &=& \sqrt {{{\hat x}^2} + {{\hat y}^2}} ,\quad S = I\left[ {1 + \tanh
\left( {{k_s}(I - {h_s})} \right)} \right] \;. \nonumber
\end{eqnarray}
As mentioned before, the desired asynchronous state of the population
corresponds to the equilibrium point $(X,Y)=(0,0)$. In
this section, we will investigate the stability of this point. The stability
analysis gives us some insight into selecting appropriate values for the
stimulation's parameters.

For simplicity of calculations, using \eqref{x_hat_y_hat}, we rewrite the
control law \eqref{control_law} as:
\begin{equation}\label{control_law_simple}
u = \Upsilon ({x_2} + \Pi {x_3}),
\end{equation}
where $\Upsilon  =  - \gamma \delta \cos \alpha$ and $\Pi  = \mu {\omega _0}\tan
\alpha$. For this new representation, we consider the case $\cos \alpha  \ne 0$.
At the points  $\alpha  = (2k - 1)\pi /2,\,k=1,2,...$ the control term is
calculated as $u =  \pm \gamma \delta \mu {\omega _0}{x_3}$.

\begin{figure}[ht]
\centerline{\includegraphics[width=1\columnwidth]{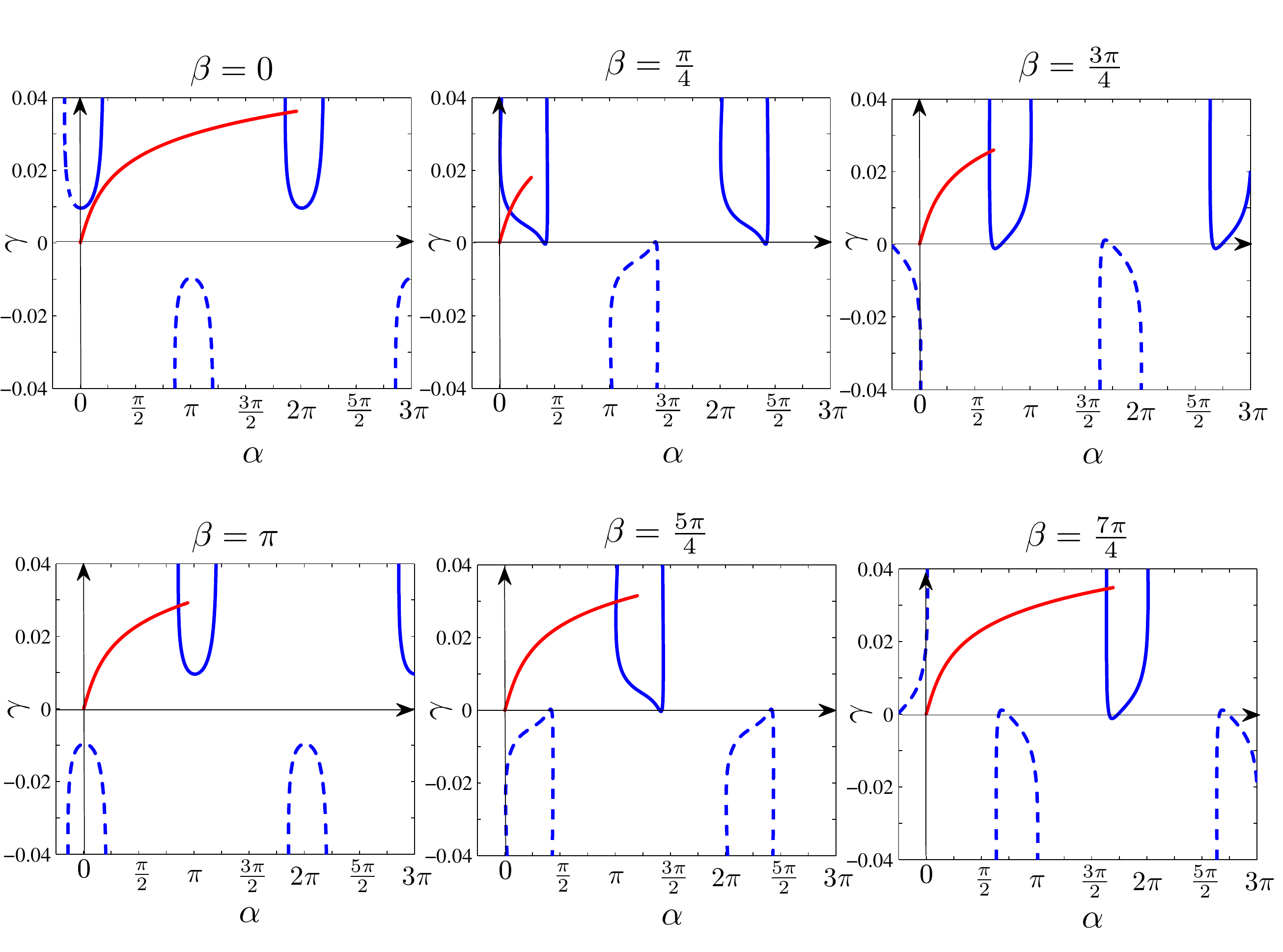}}
\caption{ (Color online)  Stability region for the controlled system (\ref{cl_1})-(\ref{cl_5}) consists of closed
sub-regions; here they are shown for different values of $\beta$. 
Only areas of the strong stability are shown here. 
The borders of the stability sub-regions in the positive quadrant are shown by solid lines. 
(For completeness, we also show by dashed lines the stability borders for negative $\alpha$, $\gamma$.) 
For each value of $\beta$, the trajectory followed by the adaptive variables $\alpha$ and $\gamma$
are plotted; eventually, they  trap in one of the stability sub-regions. }
\label{stability_domain}
\end{figure}

At the first step of analysis, we study linear stability of the only equilibrium point
$(X,Y,x_1,x_2,x_3)=(0,0,0,0,0)$ in the absence of adaptation, i.e. for  fixed values of $\alpha$ and $\gamma$.
For this aim, it is enough to consider linearization of  Eqs.~(\ref{cl_1})-(\ref{cl_5}).
The problem reduces to that of classifying the eigenvalues $\lambda$ of the 
state matrix $\mathcal A$
according to their real parts; at stability borders they are purely imaginary
$\lambda=i\Omega$. This allows one to find stability borders on the planes
$(\Upsilon,\Pi)$ or $(\alpha,\gamma)$ in a parametric form, 
as functions of parameter $\Omega$ 
(for details see Appendix \ref{ap:stab}). Based  on these steps,  
in Fig.~\ref{stability_domain} we plot the stability
region obtained for the following values of systems' parameters: $\xi=0.0048$,
$\omega_0=2\pi/32.5$, $\delta=0.3\omega_0$, $\mu=500$ (we refer to these values
in Subsection~\ref{Stuart-Landau}) and six samples of $\beta$ as 
representatives of  possible values of $\beta\in[0,2\pi)$.
It is easy to check that if the point
$({\alpha ^*},{\gamma ^*})$ belongs to the stability borders, then the points
$({\alpha ^*} \pm 2k\pi ,{\gamma ^*})$  and $({\alpha ^*} \pm (2k - 1)\pi , -
{\gamma ^*})$  belong to the stability borders as well.

At the second step of the analysis, we reformulate the linear stability results
from Lyapunov theory point
of view.
Suppose  $\mathfrak X$    is the vector of the first five variables in
Eqs.~(\ref{cl_1})-(\ref{cl_5}), i.e., 
$\mathfrak X  = {(X,Y,{x_1},{x_2},{x_3})^T}$,  where
the superscript $T$ denotes the transpose. The linearized system around $\mathfrak X=0$ can be
written as:
\begin{equation}\label{5_linear_dynamics}
\dot {\mathfrak X}  = \mathcal  A{\mathfrak X}, 
\end{equation}
where the elements of the state matrix $\mathcal A$ are functions of the
systems' parameters and variables $\alpha$ and $\gamma$. We have shown that, for
each $\beta$, there is the stability region (composed of some sub-regions) in the
$\alpha-\gamma$ plane such that if $\alpha$ and $\gamma$ are selected in this
region, the controlled system Eqs.~(\ref{cl_1})-(\ref{cl_5}) 
is linearly (locally
exponentially) stable at $\mathfrak X=0$. In other words, independently of the
value of $\beta$, there is a non-empty region $\mathcal N$ in the positive
quadrant, such that for $(\alpha,\gamma)\in{\mathcal N}$
local exponential stability  of the system \eqref{5_linear_dynamics} is
guaranteed. As is stated by the converse Lyapunov theorem (\cite{Khalil:2002} Sec. 4.7),
there exist a positive definite Lyapunov function $V_{\mathcal N}(\mathfrak X)$
($V_{\mathcal N}(\mathfrak X)>0$ and $V_{\mathcal N}(0)=0$) and a region ${{\mathcal
N}_S}$ ($ {{\mathcal
N}_S} \subset {\mathcal N}$) that satisfies the
following condition
\begin{equation}\label{derivative_V_5_states}
\dot {V}_{\mathcal N}  = \frac{{\partial {V_{\mathcal N}}}}
{{\partial{\mathfrak X}}}\dot {\mathfrak X} \leqslant  - M(\alpha ,\gamma
){\left\| \mathfrak X \right\|^2},\,\,  \forall (\alpha ,\gamma ) \in {{\mathcal
N}_S}, 
\end{equation}
for some positive function  $M(\alpha,\gamma)$ which may be the function of
other systems' parameters. Here $\left\|.\right\|$  denotes the
Euclidean 2-norm. 
Let ${M_m} = \mathop {\min }\limits_{\alpha ,\gamma  \in {{\mathcal
N}_s},\,\,\beta  \in [0,2\pi )} M(\alpha ,\gamma )$, then
Eq.~\eqref{derivative_V_5_states} is simplified to:
\begin{equation}\label{derivative_V_5_states_simple}
\dot {V}_{\mathcal N}   \leqslant  - M_m{\left\| \mathfrak X
\right\|^2},\,\,\,\,\,  \forall (\alpha ,\gamma ) \in {{\mathcal N}_S}. 
\end{equation}
We refer to \eqref{derivative_V_5_states_simple} later in this section.

At this point we add the dynamics of the adaptive 
variables $\alpha$ and $\gamma$. Consider
the new Lyapunov function $V=V_{\mathcal N}+\frac{1}
{2}({\dot \alpha ^2} + {\dot \gamma ^2})$. $V$ is  a positive function of its
arguments and is equal to zero at $\mathfrak X=0=\dot \alpha =\dot
\gamma=0$, which according to Eqs.~(\ref{switch})-(\ref{adaptive_variables_gamma})
results in $I=0$. Therefore, $V$ is a positive definite function and can be a
candidate for the Lyapunov function. In the sufficiently small vicinity of
$I=0$, the $S$ function can be approximated by the linear term $KI$ where
$K=1-\tanh(k_sh_s)$. This approximation leads to the following description for
$\alpha$ and $\gamma$:
\begin{eqnarray}
  { \dot \alpha}& =&{k_\alpha
}KI\;, \label{linear_adaptive_variables_alpha} \\
{ \dot \gamma} &=&  {{{k_{{\gamma _1}}}KI}}
{{\cosh^{-1} \left( {{{{k_{{\gamma _2}}}/{\gamma ^*}}}
{{{\omega _0}}}} \right)}},\;\label{linear_adaptive_variables_gamma}
\end{eqnarray}
where $\gamma^*$ is the steady-state value of $\gamma$.
Now, differentiating the Lyapunov function $V$ along the trajectories of the
augmented system \eqref{5_linear_dynamics},   \eqref{linear_adaptive_variables_alpha} 
and \eqref{linear_adaptive_variables_gamma} results in:
\begin{equation}\label{derivative_V}
\begin{gathered}
  \dot V = \dot{V}_{\mathcal N} + \dot \alpha \ddot \alpha  + \dot \gamma \ddot
\gamma  \hfill =  \dot{V}_{\mathcal N}  + {K^2}\left( {{{k_{{\gamma
_1}}^2}}
{{{{\cosh }^{-2}}\left( {{{{k_{{\gamma _2}}}{\gamma ^*}}}/{{{\omega _0}}}}
\right)}} + k_\alpha ^2} \right)I\dot I \hfill \\
  \,\,\,\,\, \leqslant  \dot{V}_{\mathcal N} + {K^2}(k_{{\gamma _1}}^2 +
k_\alpha ^2)I\dot I \;.
\end{gathered} 
\end{equation}
Replacing $I\dot I = {\delta ^2}({x_2}{\dot x_2} + {(\mu {\omega
_0})^2}{x_3}{\dot x_3})$ in \eqref{derivative_V} and substituting the dynamics
of $x_2$  and $x_3$ (Eqs. \eqref{cl_4} and \eqref{cl_5}), we simplify Eq.~\eqref{derivative_V} to:
\begin{equation}\label{derivative_V_simple}
\dot V\, \leqslant  \dot{V}_{\mathcal N}  + {K^2}(k_{{\gamma _1}}^2 + k_\alpha
^2){\mathfrak X^T}P\mathfrak X\;,
\end{equation}
where
$$P = \left[ {\begin{array}{*{20}{c}}
   0 & 0 & 0 & {0.5} & 0  \\
   0 & 0 & 0 & 0 & 0  \\
   0 & 0 & 0 & { - 0.5\omega _0^2} & 0  \\
   {0.5} & 0 & { - 0.5\omega _0^2} & { - \delta } & {0.5\mu \omega _0^2}  \\
   0 & 0 & 0 & {0.5\mu \omega _0^2} & { - \mu \omega _0^2}  \\
 \end{array} } \right].$$
We know that 
${\mathfrak X^T}P\mathfrak X\leqslant \lambda_{max} (P)\left\|
\mathfrak X \right\|^2$, where $\lambda_{max} (P)$ is the largest eigenvalue of
$P$. It can be easily checked that $P$ has two zero eigenvalues and at least one
positive eigenvalue. Thus, $\lambda_{max} (P)$ is non-zero and positive. Based
on these properties, Eq.~\eqref{derivative_V_simple} can be rewritten as:
\begin{equation}\label{derivative_V_simple_unified}
\dot V \leqslant  \dot{V}_{\mathcal N}  + \lambda_{max} (P){K^2}(k_{{\gamma
_1}}^2 + k_\alpha ^2)\left\| \mathfrak X \right\|^2
\end{equation}
Using Eq.~\eqref{derivative_V_5_states_simple}, in the region $\mathcal N_S$ we have:
\begin{equation}\label{derivative_V_simple_final}
\dot V \leqslant -[M_m- \lambda_{max} (P){K^2}(k_{{\gamma _1}}^2 + k_\alpha
^2)]\left\| \mathfrak X \right\|^2,\,  \forall (\alpha ,\gamma ) \in {{\mathcal
N}_S}
\end{equation}
Now, if $\lambda_{max} (P){K^2}(k_{{\gamma _1}}^2 + k_\alpha ^2)<M_m$, then
$\dot V<0,\,  \forall (\alpha ,\gamma ) \in {{\mathcal N}_S}$. This means that,
$ \left\|\mathfrak X \right\|$ and, thus, their elements decrease, provided $(\alpha ,\gamma )$ belong to 
the region  ${\mathcal N}_S$. Now  we have to decide, whether it is possible that
during this evolution $(\alpha ,\gamma )$ leave ${\mathcal N}_S$. To this end we
estimate the total shift of $(\alpha ,\gamma )$, 
assuming they enter the  region ${\mathcal N}_S$ with 
some initial values $(\alpha_0 ,\gamma_0 )$. 
As one can see from \eqref{linear_adaptive_variables_alpha} and \eqref{linear_adaptive_variables_gamma}, 
for small $k_{\alpha}$  and  $k_{\gamma_1}$ these adaptive 
parameters evolve slowly, so we can separate the time scale 
of their evolutions from the time scale of the evolution of $\mathfrak X$;
the latter is determined by $\lambda_1$,
 the closest to the imaginary axis eigenvalue of matrix
$\mathcal  A$. In this approximation, the relaxation of the auxiliary 
oscillating modes is given by $I=I_0\exp[\text{Re}(\lambda_1) t]$ 
(where one can take $\lambda_1$ in the middle of the 
region ${\mathcal N}_S$). Substituting this into 
equations \eqref{linear_adaptive_variables_alpha} and \eqref{linear_adaptive_variables_gamma} and then
integrating them we get
for the shifts of the adaptive parameters
\begin{align*}
\Delta\alpha&={k_\alpha
}KI_0|\text{Re}(\lambda_1)|^{-1}\;,\\
  \Delta \gamma & = k_{\gamma _1}KI_0
|\cosh \left( {{{{k_{{\gamma _2}}}/{\gamma ^*}}}
{{{\omega _0}}}} \right) \text{Re}(\lambda_1)|^{-1}\;.
\end{align*}
Since $K$ is exponentially small, these shifts are small as well, 
and thus the adaptive variables $(\alpha,\gamma)$ remain in 
the same region ${\mathcal N}_S$.

Adaptation parameters $k_{\alpha}$  and  $k_{\gamma_i},\,i=1,2$ play the key
role in the behavior of the controlled system. They determine the trajectory in
the $(\alpha,\gamma)$ plane. This
trajectory always starts from the origin and terminates in one of the stability
sub-regions. As it can be seen from Fig.~\ref{stability_domain}, for each
$\beta$, the stability region consists of  periodically arranged sub-regions. 
Some of them are
not accessible by $\alpha$ and $\gamma$ because of their increasing dynamics, 
which keep them in the positive quadrant of the $\mathbb R^2$
space. In Fig.~\ref{stability_domain} the possible stability sub-regions are
shown with the solid lines. For the fixed $\beta$, each of these stability
sub-regions have the potential to be the terminal point for the adaptive variables. However,
the values of the adaptation parameters and also of $S$ determine, which one will be 
selected.

\section{Applying the adaptive stimulation to ensembles of coupled oscillators}
\label{simulation_results}
In this section, we apply the proposed stimulation to different ensembles of
coupled oscillators. We start by an ensemble of Stuart-Landau oscillators in order to
compare the simulation results with the theory. Then, in subsections~\ref{van_der_pol} and ~\ref{hindmarsh}, 
we perform simulations which reveal the efficiency of the stimulation in desynchronizing ensembles of more
complex and realistic oscillators. For better readability, we present the 
details of parameters  sets in Appendix~\ref{ap}.

\subsection{Stuart-Landau oscillators}
\label{Stuart-Landau}

 A simple Stuart-Landau model reflects the key properties of a
self-sustained oscillator. Hence, as the first example, we consider an ensemble
of $N=1000$ all-to-all coupled and non-identical Stuart-Landau oscillators as:
\begin{equation}\label{controlled_Stuart-Landau_ensemble}
\begin{array}{l}
 {{\dot x}_i} = {a}{x_i} - {\omega _i}{y_i} - {x_i}(x_i^2 + y_i^2) + CX +   u\cos \beta,\\ 
 {{\dot y}_i} = {\omega _i}{x_i} + {a}{y_i} - {y_i}(x_i^2 + y_i^2) + CY + u \sin \beta,\\ 
 \end{array}
\end{equation}
where $i=1,2,...,N$. Here $C$ is the strength of the coupling via the mean fields $X={N^{ -
1}}\sum\limits_i {{x_i}}$  and   $Y={N^{ - 1}}\sum\limits_i
{{y_i}}$   and $u$ is the control term  based
on the measured signal $m=X$. 

Numerical simulations of a non-controlled population show that dynamics of
Eqs.~\eqref{controlled_Stuart-Landau_ensemble} with the parameters' 
values as in Appendix~\ref{ap:sl} is quite well described by the 
macroscopic model
\eqref{controlled_complex_macroscopic_model}  with $\xi=0.0048$ and
$\omega_0=2\pi/32.5$.  So, without considering the
adaptation mechanism, the borders of the stability region of the controlled
system can be analytically approximated by ones plotted in
Fig.~\ref{stability_domain}. Now, we want to investigate whether the adaptive
mechanism can force $\alpha$ and $\gamma$ to trap in one of the stability
sub-regions. To this aim, we select $k_\alpha=0.003$,  $k_{\gamma_1}=0.0001$
and $k_{\gamma_2}=20$. Then, we simulated the controlled system
\eqref{controlled_Stuart-Landau_ensemble} for several values of $\beta$, as shown 
in sub-panels of Fig.~\ref{stability_domain}. In the $\alpha-\gamma$ plain, we
plot the trajectories tracked by the adaptive variables. As can be seen, for this
choice of adaptation parameters all of the $(\alpha,\gamma)$-trajectories move
toward the nearest stability sub-region (except for $\beta=0$).

\begin{figure}[h]
\centerline{\includegraphics[width=.95\columnwidth]{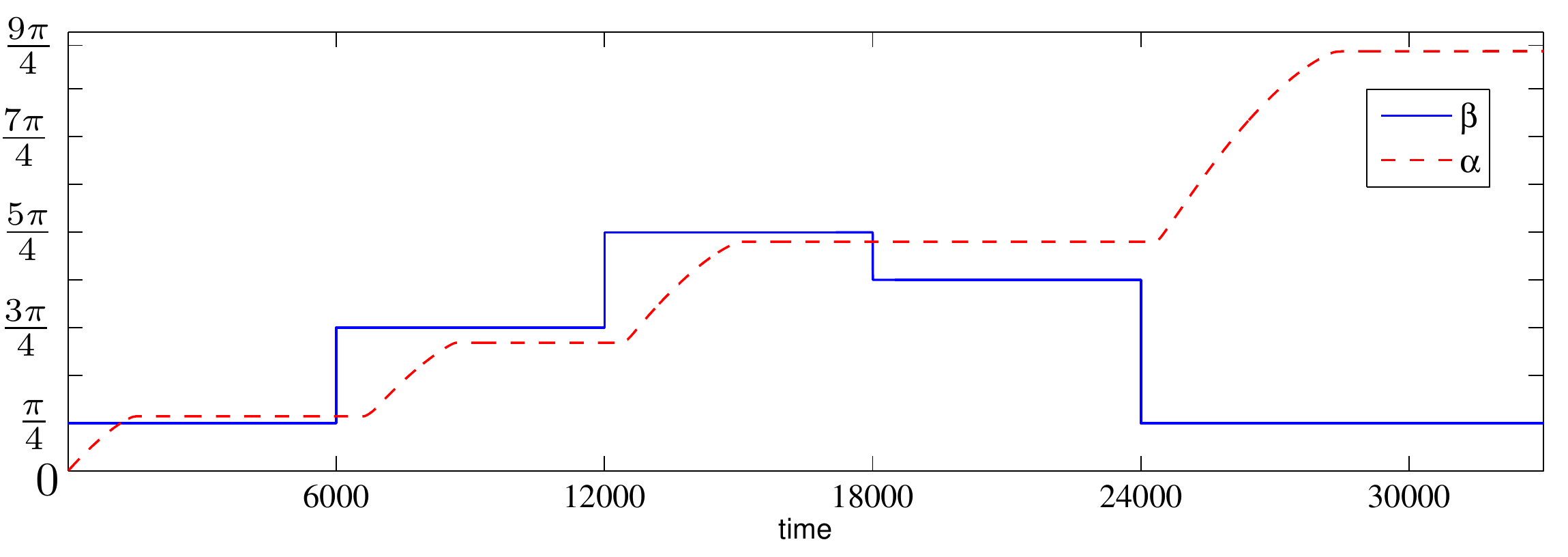}}
\caption{ (Color online) Time course of $\beta$  in
Eqs.~\eqref{controlled_Stuart-Landau_ensemble}  (blue solid line) and the 
corresponding adaptation of $\alpha$ (red dashed line). For parameters, see Appendix \ref{ap:sl}.}
\label{alpha_beta_tracking}
\end{figure}

In the next simulation, we make $\beta$ in
\eqref{controlled_Stuart-Landau_ensemble}  time-dependent, as shown in 
Fig.~\ref{alpha_beta_tracking}. We plot there also the corresponding
evolution of $\alpha$. Notice, that in our control scheme 
parameter $\alpha$ cannot decrease. 
One can see that when $\beta$ increases, $\alpha$ 
approximately follows $\beta$. When $\beta$ decreases, first $\alpha$ avoids to
evolve and the controller tries to preserve the obtained asynchronous state
based on the previous value of $\alpha$.
However, larger change of $\beta$ breaks the controller inertia and forces
$\alpha$ to adapt by increasing by $\approx 2\pi$ to reach  
the next stability sub-region. 

\subsection{Bonhoeffer-van der Pol  oscillators}
\label{van_der_pol}
Self-sustained Bonhoeffer-van der Pol oscillator,  a system close to the FitzHugh-Nagumo neuron 
model, oscillates around a non-zero equilibrium points. A  population of $N=1000$ coupled
oscillators under the stimulation $u$ is described by:
\begin{equation}\label{controlled_van_der_pol_ensemble}
\begin{array}{l}
 {{\dot x}_i} = {x_i} - x_i^3/3 - {y_i} + {I_i} + CX + u  \cos \psi,\\ 
 {{\dot y}_i} = 0.1({x_i} - 08{y_i} + 07) +  u \sin \psi. \\ 
 \end{array}
\end{equation}
\begin{figure}[h]
\centerline{\includegraphics[width=.95\columnwidth]{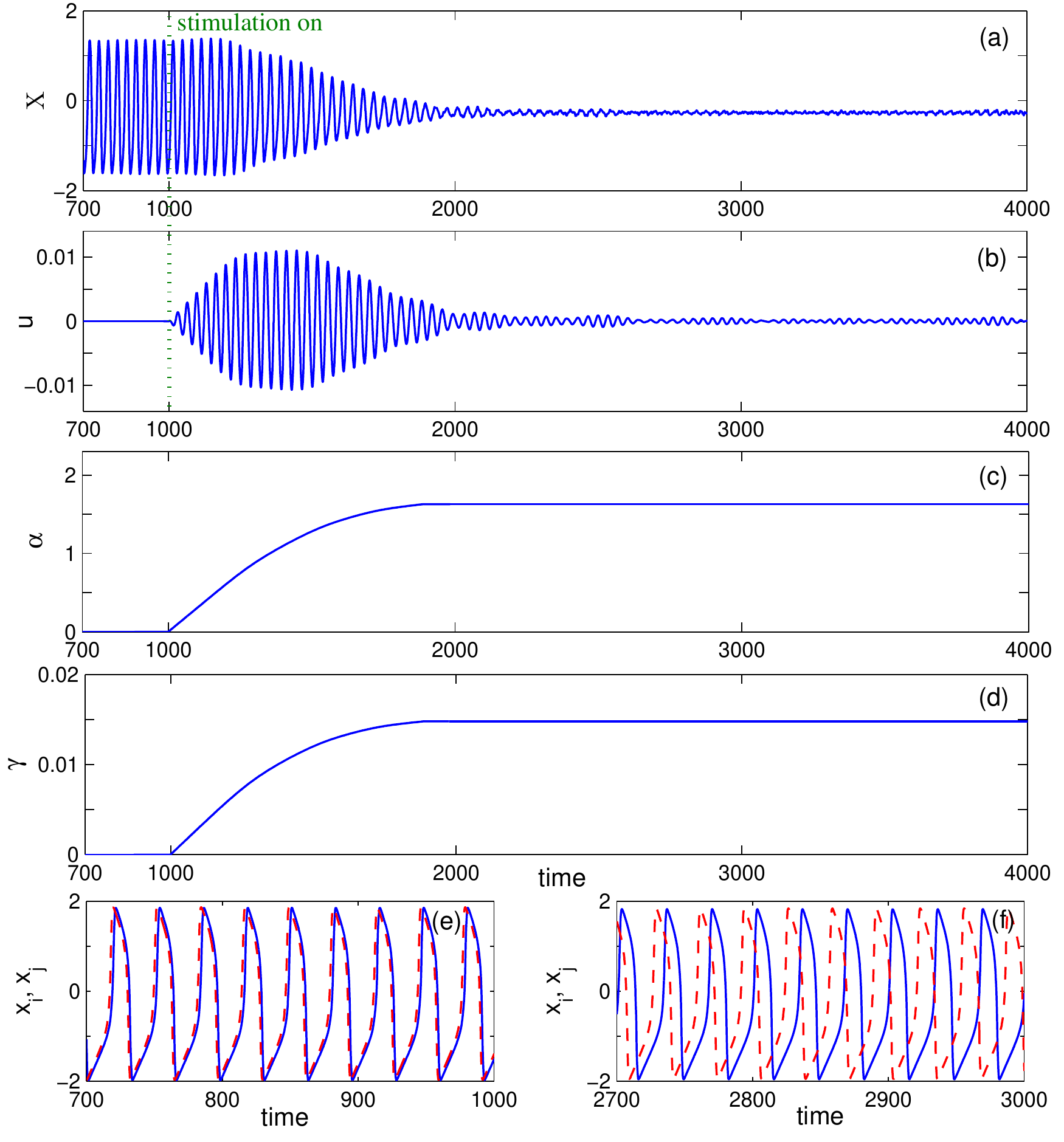}}
\caption{ (Color online) Desynchronization in an ensemble of Bonhoeffer-van der
Pol  oscillators \eqref{controlled_van_der_pol_ensemble} with $\psi=\pi/2$. 
(a) The mean field
$X$, (b) the control signal $u$ and (c),(d) the adaptive variables vs time. (e,f)
Synchronous and asynchronous dynamics of two arbitrary chosen oscillators in the ensemble, before and after applying the stimulation, 
respectively. For parameters see
Appendix~\ref{ap:vdp}.}
\label{van_der_pol_1_beta}
\end{figure}
 In this example, the impact of the stimulation $u$ on the
oscillators is again distributed between the $x$ and $y$ equations 
according to the value of $\psi$. As mentioned in \cite{Tukhlina_et_al:2007}, 
the parameter $\psi$ is
related but not equal to the phase shift parameter $\beta$ in
Eq. \eqref{controlled_complex_macroscopic_model}.
The measured signal $m=X$ can be considered as contaminated by some intrinsic noise due to the
finite ensemble size. Next, it has a non-zero average.  
Simulation results are depicted in Fig.~\ref{van_der_pol_1_beta}.
The stimulation is applied at  $t=1000$. Before that, the oscillators are
synchronized which is reflected in the large amplitude of the mean field $X$.
When the stimulation is switched on,  it smoothly affects the
synchronized oscillators. As a result, the mean field gradually vanishes at the
level of the induced noise which in turn results in vanishing stimulation. When
the asynchronous state is achieved, the vanishing simulation maintains it.  
The evolution of
$\alpha$ and $\gamma$ is shown in Fig.~\ref{van_der_pol_1_beta}(c) and (d),
respectively. Finally, the behaviors of two arbitrary oscillators in the
ensemble (Fig.~\ref{van_der_pol_1_beta}(e) and (f)), 
verify the capability of the stimulation in destroying the synchronous
state  without destroying individual oscillations.
\begin{figure}[t]
\centerline{\includegraphics[width=1\columnwidth]{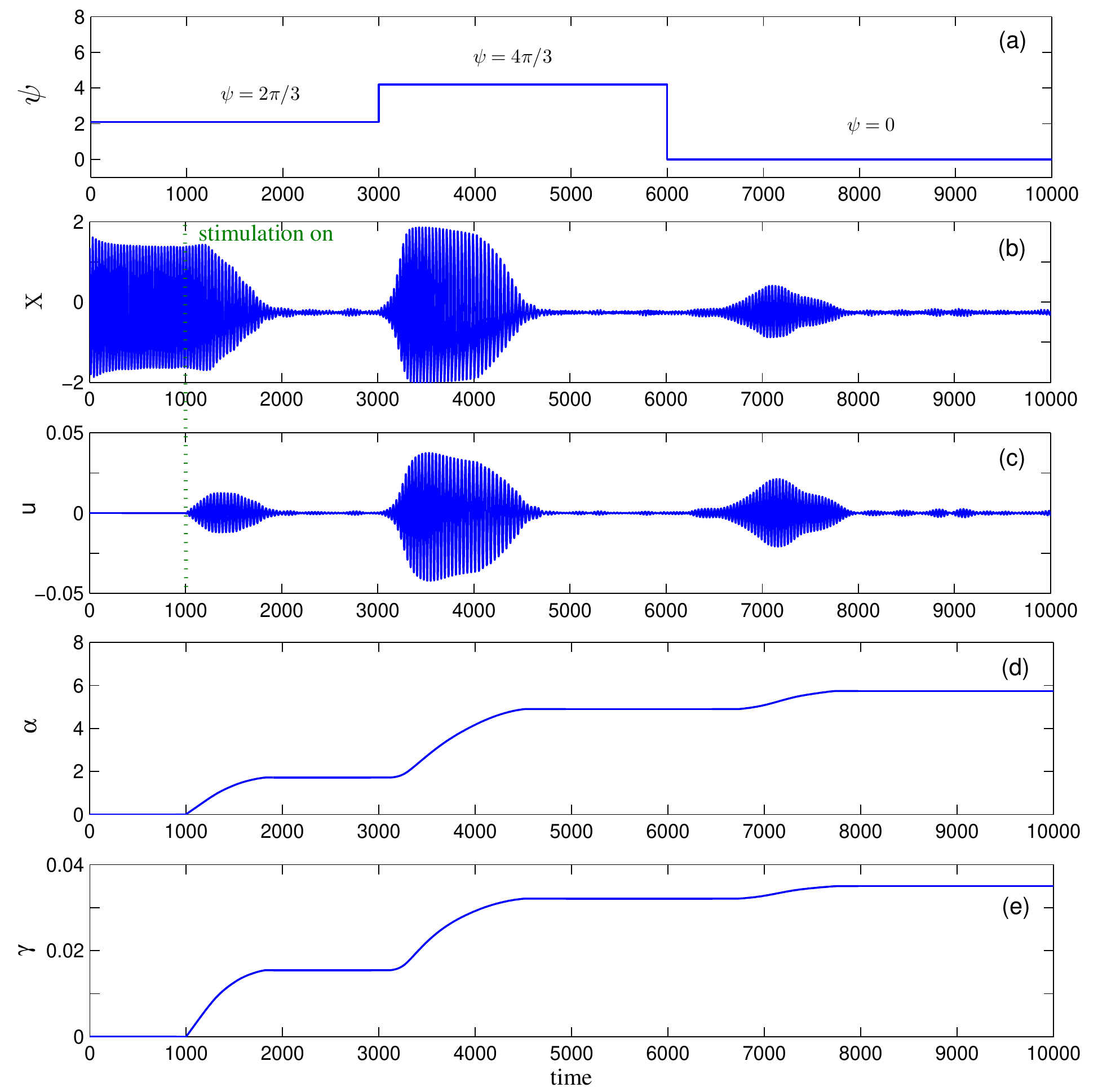}}
\caption{Suppressing synchrony in an ensemble of Bonhoeffer-van der Pol 
oscillators in the presence of the time-variant phase shift parameter $\psi$.
When $\psi$  changes in time  (a), the adaptive variables start to evolve (d)
and (e), so that the stimulation (c) adapts itself to the new values of
$\beta$ and stabilizes the mean field (b). 
Parameters are the same as in Fig.~\ref{van_der_pol_1_beta}.}
\label{van_der_pol_3_beta}
\end{figure}

To show the capability  of
the proposed stimulation in coping with the case of 
time-dependent system parameters, 
in the next simulations
(Fig.~\ref{van_der_pol_3_beta}) we vary $\psi$ as shown in panel (a). 
The controllers' parameters are as
before. As one can see, for $\psi=2\pi/3$ the stimulation quenches
the mean field. However, as $\psi$ changes to $\psi=4\pi/3$, the mean field
start growing and the feedback mechanism starts adapting the stimulation
to the new situation. $\alpha$
and $\gamma$ begin to evolve  and finally they
settle down to the new values at which the stimulation breaks the synchrony.
Depending on the vale of $\psi$, the new value of the adaptive variables may be
found quickly or slowly, which corresponds to fast or slow damping of the mean
field, respectively (compare the damping time for $\psi=4\pi/3$ and $\psi=0$).

Noteworthy, if the already controlled regime needs to be adapted,
mostly fast are variations of $\alpha$, while variations of $\gamma$
are slowed down by the denominator in the $\dot \gamma$ equation;
with this we minimize the amplitude of achieved feedback control. 


\subsection{Hindmarsh-Rose neurons with synaptic coupling}
\label{hindmarsh}

Since the main purpose of our technique is to 
suppress neuronal synchrony, in this subsection we analyse a more realistic 
model. We consider an
ensemble of Hindmarsh-Rose neurons and discuss in more detail the 
measurement of the collective activity and the coupling between the units.
The model of  $N$ all-to-all coupled units reads:
\begin{equation}\label{controlled_hindmarsh_rose_ensemble}
\begin{aligned}
 {{\dot x}_i} &= 3x_i^2 - x_i^3 + {y_i} - {z_i} + {I_i} - \frac{C}{{N - 1}}({x_i} + {V_c})
  \sum\limits_{j \ne i}^N {{{\left[ {1 + {e^{\left( {\frac{{{x_i} - {x_0}}}{\eta }} \right)}}} \right]}^{ - 1}} + u}\;,  \\ 
 {{\dot y}_i} &=  - 5x_i^2 - {y_i} + 1\;, \\ 
 {{\dot z}_i} &= r[\nu ({x_i} - \chi ) - {z_i}]\;. \\ 
 \end{aligned}
\end{equation}
Depending on the value of the parameters, the units 
show spiking or bursting. 
Neuronal oscillators  in \eqref{controlled_hindmarsh_rose_ensemble}
interact via the synaptic connections, described with the inverse exponential term. 
This type of coupling plays an
important role in  large networks of neurons where even spatially distant
neurons can be linked by long axons.

\begin{figure}[h]
\centerline{\includegraphics[width=1\columnwidth]{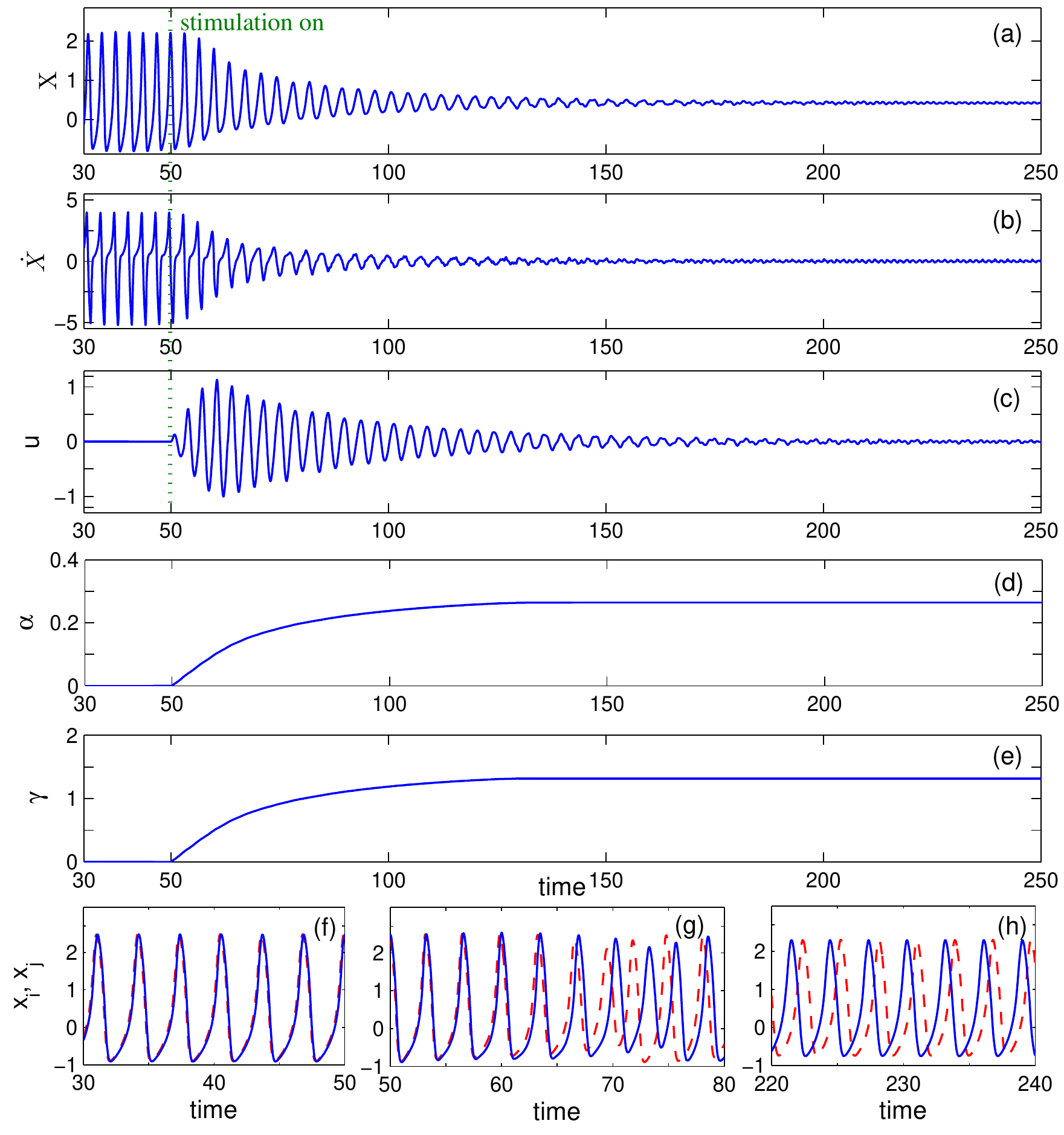}}
\caption{ (Color online) Suppression of synchrony in an ensemble of synaptically
coupled spiking Hindmarsh-Rose neurons. (a)-(c) The mean field, its derivative
(which is used as a measured signal) and the control signal $u$, respectively.
Time courses of the adaptive variables (d) and (e).  The behavior of
two arbitrary chosen oscillators in the ensemble, before (f), exactly after (g), and  some
time after applying the stimulation (h).  For parameters, see
Appendix \ref{ap:shr}.}
\label{hindmarsh_spike}
\end{figure}

In our model, the stimulation is described by an additional external current,
common for all neurons; it enters equations for $x$. 
Following \cite{Tukhlina_et_al:2007}, we assume, that 
the measured signal used as an input to the feedback loop is proportional to 
the derivative of the mean field, $m = \dot X = {N^{ - 1}}\sum\limits_i
{{{\dot x}_i}} $.
The results of applying
the adaptive stimulation to $N=200$ coupled
oscillators \eqref{controlled_hindmarsh_rose_ensemble} in a spiking regime
are shown in Fig.~\ref{hindmarsh_spike}.
Figures~\ref{hindmarsh_spike}(a),(b) demonstrate that although the
stimulation bases on $\dot X$ instead of $X$, it retains its capability 
in desynchronizing the  population. Figures~\ref{hindmarsh_spike}(f)-(h) 
show the behaviors of two arbitrary chosen oscillators before, exactly after, 
and  some time after applying the stimulation, respectively. 
Comparison of these figures verifies that the
stimulation breaks the synchrony without any undesired effect on the 
individual oscillators. The reason is that the stimulated oscillators 
are smoothly and gradually affected by the controller.

\begin{figure}[t]
\centerline{\includegraphics[width=1\columnwidth]{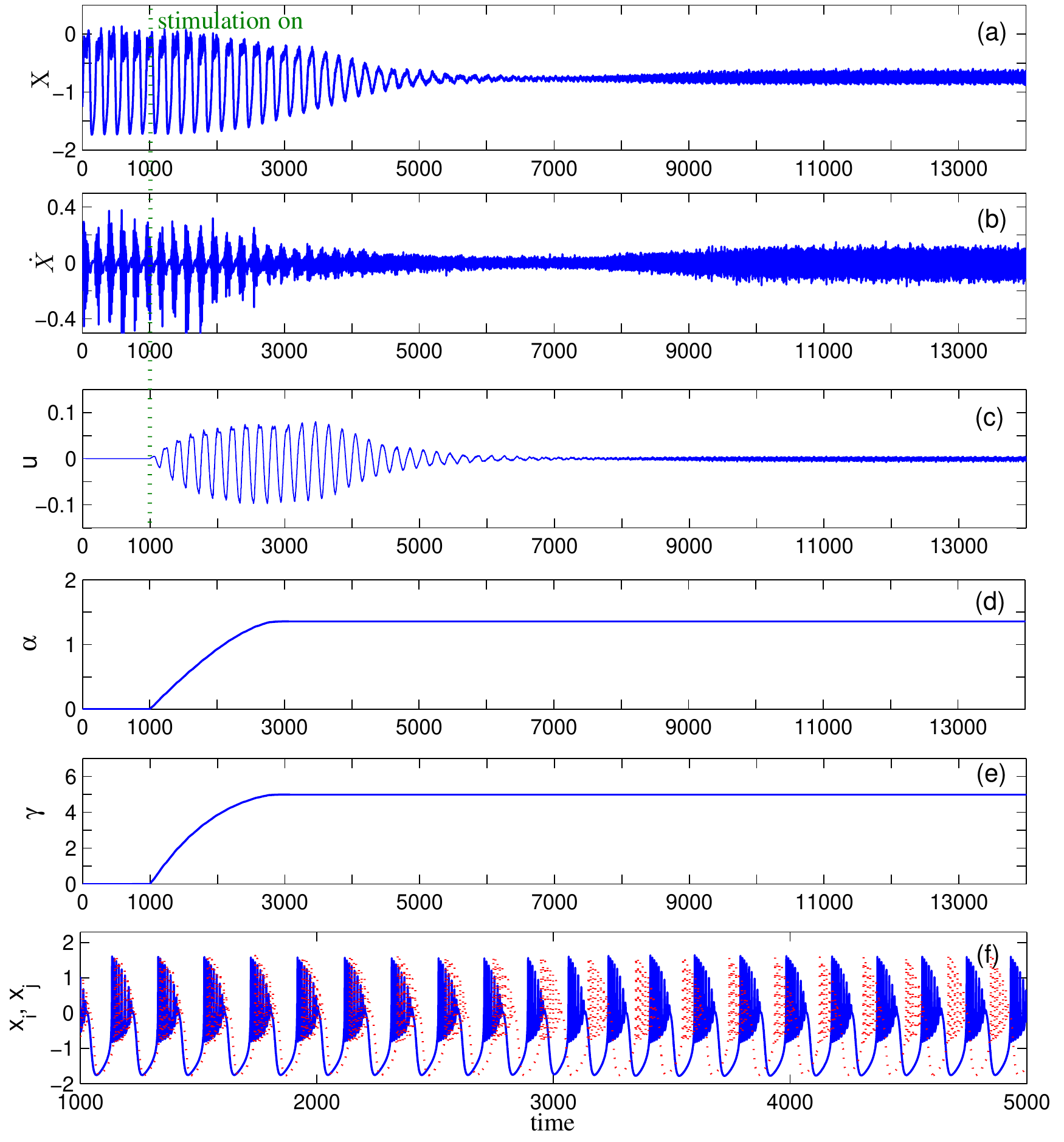}}
\caption{ (Color online) Suppression of synchrony in an ensemble of synaptically
coupled bursting Hindmarsh-Rose neurons. (a)-(c) The mean field, its derivative
(which is used as a measured signal) and the control signal $u$, respectively.
(d) and (e) Time courses of the adaptive variables. (f) Transition from 
synchrony to asynchrony illustrated with two arbitrary
chosen elements of the ensemble.  For parameters, see
Appendix~\ref{ap:bhr}. }
\label{hindmarsh_burst}
\end{figure}

  Next, we consider the case of chaotic bursting, when generation of action 
  potentials alternates with the epochs of quiescence, so that the 
  oscillations can be characterized by two time scales. 
  The simulation results
  are shown in Fig.~\ref{hindmarsh_burst}.

One important feature in an ensemble of chaotic bursting oscillators is 
that synchronization occurs on the slower time scale, i.e., different neurons 
burst nearly at the same time, whereas the spiking within the burst 
is not synchronous,
and therefore is to a large extent averaged out in the mean field. However, due
to correlations in spiking, some high frequency fluctuations remain in the mean 
field. Besides these fluctuations, a low frequency modulation of the mean field
is also observed in. 
In Fig.~\ref{hindmarsh_burst}(b), fluctuations and modulation can be
seen in the uncontrolled ensemble (i.e., $t<1000$). 
Prior to application of the stimulation, the amplitude of the measured
signal $\dot X$ is smaller than it was in the case of spiking neurons 
(cf. Fig.~\ref{hindmarsh_burst}(b) and Fig.~\ref{hindmarsh_spike}(b)). 
This means that $I$ and consequently $S$ are smaller as well. 
If we select the same adaptive parameters as for the
spiking neurons, the adaptive variables $\alpha$ and $\gamma$ vary very slowly 
and desynchronization takes quite a large time.
Therefore,  we select larger values for the adaptation parameters $k_\alpha$ and
$k_{\gamma_1}$ to speed up the dynamics of $\alpha$ and $\gamma$. In addition,
$h_s$ is taken smaller to be in harmony with the switch input $I$.
When the stimulation is switched on, it suppresses the observed periodic
components in $X$. 
Figure~\ref{hindmarsh_burst}(f) reveals that  impact of the stimulation is smooth
and two arbitrary chosen neurons gradually desynchronize.

\clearpage
\section{Conclusion}
\label{conclusion}
In this paper we have suggested a simple adaptive method for 
achieving desynchronization
in populations of oscillators via a vanishing feedback stimulation. 
The adaptation is required because a macroscopic description of a globally coupled ensemble
in terms of the model equation \eqref{controlled_complex_macroscopic_model} requires knowledge 
of three parameters. One of them, the oscillation frequency, can be easily determined from the data, 
while the  instability of the equilibrium $\xi$ and the phase parameter $\beta$ cannot. 
Therefore, two parameters, characterizing the feedback loop, namely its strength and the phase shift,
cannot be determined \textit{a priori} and should be either found by trial or by an automated adaptation 
algorithm. 

We have shown that, introducing an adaptive adjustment  of two parameters in the feedback
loop, it is possible to overcome the uncertainty in the dynamics of the mean field, varying
 the additional phase parameter and gradually increasing the amplification of the feedback loop
in such a way, 
 that a robust asynchronous state is achieved and maintained by a vanishingly small stimulation.
Moreover, sudden or smooth variations of this phase shift are successfully 
followed by the adaptation, so that after some transients asynchronous state re-establishes.

Along with the linear analysis of the scheme, we demonstrated its feasibility on
a range of models of coupled oscillators. The most nontrivial of them are 
populations of spiking and bursting neurons, described by the Hindmarsh-Rose model.
We also discussed possible restrictions on the adaptation 
parameters, although a more detailed consideration is needed in each case where the 
characteristic time scales and degree of non-stationarity of the underlying system are available.

Our research was motivated by recent studies related to neuroscience. However, the 
formulation of the control problem is quite general and is applicable to other situations 
where desynchronization of a system with unknown parameters is desirable.


\begin{acknowledgments}
We thank A. Pogromsky for helpful discussions. 
G. M. acknowledges financial support from the German Academic Exchange Service (DAAD).
 \end{acknowledgments}

\appendix
\section{ Computing the stability domain }
\label{ap:stab}
The characteristic equation of the first five equations in \eqref{closed_loop}
reads:
\begin{equation}\label{characteristic_equation}
\begin{gathered}
  {\lambda ^5}\mu  + {\lambda ^4}(1 + \delta \mu  - 2\xi \mu ) + {\lambda
^3}({\xi ^2}\mu  - \Upsilon \mu \cos \beta  + 2\omega _0^2\mu  \hfill \\
  \,\,\, - 2\xi \delta \mu  + \delta  - 2\xi ) + {\lambda ^2}(\omega _0^2\delta
\mu  - \Upsilon \cos \beta  - \Upsilon \Pi \cos \beta  \hfill \\
  \,\,\, - 2\xi \omega _0^2\mu  - 2\xi \delta  + \Upsilon \xi \mu \cos \beta  +
{\xi ^2}\delta \mu  + \Upsilon {\omega _0}\mu \sin \beta  + 2\omega _0^2 \hfill
\\
  \,\,\, + {\xi ^2}) + \lambda (\Upsilon {\omega _0}\sin \beta  + \Upsilon \xi
\cos \beta  + \Upsilon \Pi {\omega _0}\sin \beta  + {\xi ^2}\omega _0^2\mu 
\hfill \\
  \,\,\, + \omega _0^4\mu  + \omega _0^2\delta  + {\xi ^2}\delta  - 2\xi \omega
_0^2 + \Upsilon \Pi \xi \cos \beta ) + {\xi ^2}\omega _0^2 + \omega _0^4 \hfill
\\
  \,\,\,\, = 0 \hfill \\ 
\end{gathered} 
\end{equation}

The stability domain in the $\Pi-\Upsilon$ plane or, equivalently, in the
$\alpha-\gamma$ plane corresponds to the condition $\mbox{Re}(\lambda)<0$.  Thus, 
$\mbox{Re}(\lambda)=0$ gives the border of the stability region. Therefore, taking
$\lambda=i\Omega$ in the \eqref{characteristic_equation} and separating real and
imaginary parts, we obtain:
\begin{subequations}\label{borders}
\begin{align}
\begin{gathered}
  {\Omega ^4}(1 + \delta \mu  - 2\xi \mu ) + {\Omega ^2}[\Upsilon \cos \beta (1
- \xi \mu  + \Pi ) \hfill \\
 \,\,\,\,\,\,\,\, - \Upsilon {\omega _0}\mu \sin \beta  + 2\xi \omega _0^2\mu  +
2\xi \delta  - \omega _0^2\delta \mu  - {\xi ^2}\delta \mu  - 2\omega _0^2
\hfill \\
 \,\,\,\,\,\,\,\, - {\xi ^2}] + {\xi ^2}\omega _0^2 + \omega _0^4 = 0, \hfill
\\ 
\end{gathered} \label{border_real}
 \\ 
\begin{gathered}
  \Omega \{ {\Omega ^4}\mu  + {\Omega ^2}[\mu (\Upsilon \cos \beta  - {\xi ^2} -
2\omega _0^2 + 2\xi \delta ) - \delta  + 2\xi ] \hfill \\
  \,\,\,\,\,\, + \Upsilon (1 + \Pi )({\omega _0}\sin \beta  + \xi \cos \beta ) +
\omega _0^2({\xi ^2}\mu  + \delta  - 2\xi ) \hfill \\
  \,\,\,\,\,\,\,\,\,\,\, + \omega _0^4\mu  + {\xi ^2}\delta \}  = 0. \hfill \\ 
\end{gathered}  \label{border_imaginary}
\end{align}
\end{subequations} 

Since $\Omega=0$ provides no solution, we divide \eqref{border_imaginary} by
$\Omega  \ne 0$. Now, parameters $\Upsilon$ and $\Pi$ can be extracted by
solving \eqref{border_real} and \eqref{border_imaginary} as:
\begin{equation}\label{border_parameters}
\Pi  = {{{T_1}}}/
{{{T_2}}},\,\,\,\,\,\,\,\Upsilon  = {{{T_3}}}/{{{T_4}}},
\end{equation}
where 
\begin{equation}\label{border_parameters_elements}
\begin{gathered}
  {T_1} = {\omega _0}\sin \beta [{\Omega ^6}{\mu ^2} + {\Omega ^4}(1 - 2\omega
_0^2{\mu ^2} + 2{\mu ^2}\xi \delta  - {\mu ^2}{\xi ^2}) \hfill \\
  \,\,\,\,\,\,\,\, + \,{\Omega ^2}(\omega _0^2{\mu ^2}{\xi ^2} - 2\omega _0^2 +
\omega _0^4{\mu ^2} + 2\xi \delta  - {\xi ^2}) + {\xi ^2}\omega _0^2 + \omega
_0^4] \hfill \\
  \,\,\,\,\,\,\,\, + \cos \beta [{\Omega ^6}(\delta {\mu ^2} - \xi {\mu ^2}) +
{\Omega ^4}(\delta  - \xi  - {\xi ^3}{\mu ^2} + {\xi ^2}{\mu ^2}\delta  \hfill
\\
  \,\,\,\,\,\,\,\, - \omega _0^2\delta {\mu ^2}) + {\Omega ^2}({\xi ^2}\delta  -
\omega _0^2\delta  - {\xi ^3} + \xi {\mu ^2}\omega _0^4 + {\xi ^3}{\mu ^2}\omega
_0^2) \hfill \\
  \,\,\,\,\,\,\,\, + \,{\xi ^3}\omega _0^2 + \omega _0^4\xi ] \;,\hfill \\
  {T_2} = {\omega _0}\sin \beta [{\Omega ^4}(2\xi \mu  - \delta \mu  - 1) +
{\Omega ^2}({\xi ^2}\delta \mu  - 2\xi \omega _0^2\mu  - 2\xi \delta  \hfill \\
  \,\,\,\,\,\,\,\,\,\, + {\xi ^2} + \omega _0^2 + \omega _0^2\delta \mu ) - {\xi
^2}\omega _0^2 - \omega _0^4] + \cos \beta [{\Omega ^6}\mu  + {\Omega ^4}(\delta
\xi \mu  \hfill \\
  \,\,\,\,\,\,\,\,\,\, + {\xi ^2}\mu  - 2\omega _0^2\mu  + \xi  - \delta ) +
{\Omega ^2}(\omega _0^4\mu  - {\xi ^2}\omega _0^2\mu  + \omega _0^2\xi \mu
\delta  \hfill \\
  \,\,\,\,\,\,\,\,\,\, + {\xi ^3}\delta \mu  - {\xi ^2}\delta  + \omega
_0^2\delta  + {\xi ^3}) - {\xi ^3}\omega _0^2 - \omega _0^4\xi ]\;, \hfill \\
  {T_3} = \cos \beta [ - {\Omega ^6}\mu  + {\Omega ^4}(2\omega _0^2\mu  - {\xi
^2}\mu  - \mu \xi \delta  + \delta  - \xi ) \hfill \\
  \,\,\,\,\,\,\,\,\, + {\Omega ^2}(\omega _0^2{\xi ^2}\mu  - {\xi ^3} - \omega
_0^4\mu  - {\xi ^3}\delta \mu  - \omega _0^2\mu \xi \delta  + {\xi ^2}\delta  -
\omega _0^2\delta ) \hfill \\
  \,\,\,\,\,\,\,\,\, + {\xi ^3}\omega _0^2 + \omega _0^4\xi ] + {\omega _0}\sin
\beta [{\Omega ^4}(1 - 2\mu \xi  + \delta \mu ) + {\Omega ^2}(2\xi \delta 
\hfill \\
  \,\,\,\,\,\,\,\,\, + 2\xi \omega _0^2\mu  - \omega _0^2\delta \mu  - {\xi
^2}\delta \mu  - {\xi ^2} - 2\omega _0^2) + \omega _0^4 + {\xi ^2}\omega _0^2]
\;, \hfill \\
  {T_4} = \mu {\Omega ^2}[\xi {\omega _0}\sin 2\beta  + \omega _0^2 + {\cos
^2}\beta ({\Omega ^2} + {\xi ^2} - \omega _0^2)] \;.\hfill \\ 
\end{gathered} 
\end{equation}
Now, for each value of $\Omega$, using \eqref{control_law_simple}, we compute:
\begin{equation}\label{alpha_gamma_on_border}
\alpha  =\arctan \left( {{{{T_1}}}
/{{{T_2}\mu {\omega _0}}}} \right),\,\,\,\,\,\,\gamma  =  - \,{{{T_3}}}/
{{{T_4}\delta \cos \alpha }}
\end{equation}

\section{Parameters of oscillators ensembles}
\label{ap}
\subsection{Stuart-Landau oscillators}
\label{ap:sl}

For the Stuart-Landau oscillators \eqref{controlled_Stuart-Landau_ensemble}, $a=0.01$ and
$\omega_i$ are selected from a Gaussian distribution with the mean value
$w_0=2\pi/32.5$ and rms value $0.001$; $C=0.008$. With these
values of parameters,  the mean
field's dynamics can be approximated by
Eq.\eqref{uncontrolled_complex_macroscopic_model} with $\xi=0.0048$ and
$\omega_0=2\pi/32.5$. For the bandpass filter
and the integrator we select $\delta=0.3\omega_0$ and $\mu=500$. These values
are exactly the same as the ones used for the simulations shown in
Fig.~\ref{stability_domain}. The switch and the adaptation parameters  are selected  as  $h_s=0.05$, $k_s=200$ and $k_\alpha=0.003$, $k_{\gamma_1}=0.0001$, $k_{\gamma_2}=20$, respectively.

\subsection{Bonhoeffer-Van der Pol oscillators}
\label{ap:vdp}

In \eqref{controlled_van_der_pol_ensemble}, $I_i$ are selected from a Gaussian distribution
with the mean value $0.6$ and rms value $0.1$.  
When $u=0$ and $C<0.018$ the mean field $X$ shows small irregular
fluctuations around $X_0=-0.26$. These fluctuations are due to the finite size
of the population. Since the equilibrium point of the individual oscillators is
not at zero, the mean field has a constant term. For $C>0.018$ the population
synchronizes which leads to large oscillation of $X$. 
For simulation we take $N=1000$ and the following 
parameters' values: $C=0.03$, 
$\omega_0=2\pi/32.5$, $\delta=0.3\omega_0$,
$h_s=0.2$, $k_s=500$, $k_\alpha=0.001$, $k_{\gamma_1}=10^{-5}$, and 
$k_{\gamma_2}=10$.

\subsection{Spiking Hindmarsh-Rose neurons}
\label{ap:shr}

In the Hindmarsh-Rose neuron model \eqref{controlled_hindmarsh_rose_ensemble}, 
we set the parameters' values as:
$r=0.006$, $\nu=1$, and $\chi=-1.56$. The coupling strength is 
$C=0.4$ which results in synchronous oscillations in the absence of the
stimulation $u$. Other parameters of synapses are $\eta=0.01$, $x_0=0.85$, and
inverse potential $V_c=1.4$. The external current $I_i$ is taken as
$I_i=6+\sigma$, where $\sigma$ is Gaussian distributed with zero mean and $0.1$
rms value.

Without the stimulation, the synchronized ensemble shows oscillations with the
average frequency $\omega_0=2\pi/3.2$. Again, the parameters of the filter and
the integrator are $\delta=0.3\omega_0$ and $\mu=500$. 
Finally, he parameters' values of the
the adaptive stimulation are: $h_s=0.15$, $k_s=200$,
$k_\alpha=0.002$, $k_{\gamma_1}=0.01$, and $k_{\gamma_2}=0.01$.

\subsection{Bursting Hindmarsh-Rose neurons}
\label{ap:bhr}

The chaotic bursting oscillation in \eqref{controlled_hindmarsh_rose_ensemble} 
are obtained by taking $\nu=4$, $\chi=-1.6$, and $I_i=3.2$. Parameters of the coupling 
are kept as before. The coupling strength $C=0.2$. 
The average frequency of the
mean field is $\omega_0=2\pi/176$. The stimulation's parameters are as follows:
switch parameters $h_s=0.01$, $k_s=500$, adaptation parameters $k_\alpha=0.02$,
$k_{\gamma_1}=0.1$, and $k_{\gamma_2}=0.01$.


\begin{thebibliography}{10}%
\makeatletter
\providecommand \@ifxundefined [1]{%
 \ifx #1\undefined \expandafter \@firstoftwo
 \else \expandafter \@secondoftwo
\fi
}%
\providecommand \@ifnum [1]{%
 \ifnum #1\expandafter \@firstoftwo
 \else \expandafter \@secondoftwo
\fi
}%
\providecommand \enquote [1]{``#1''}%
\providecommand \bibnamefont  [1]{#1}%
\providecommand \bibfnamefont [1]{#1}%
\providecommand \citenamefont [1]{#1}%
\providecommand\href[0]{\@sanitize\@href}%
\providecommand\@href[1]{\endgroup\@@startlink{#1}\endgroup\@@href}%
\providecommand\@@href[1]{#1\@@endlink}%
\providecommand \@sanitize [0]{\begingroup\catcode`\&12\catcode`\#12\relax}%
\@ifxundefined \pdfoutput {\@firstoftwo}{%
 \@ifnum{\z@=\pdfoutput}{\@firstoftwo}{\@secondoftwo}%
}{%
 \providecommand\@@startlink[1]{\leavevmode}%
 \providecommand\@@endlink[0]{}%
}{%
 \providecommand\@@startlink[1]{%
  \leavevmode
  \pdfstartlink
   attr{/Border[0 0 1 ]/H/I/C[0 1 1]}%
   user{/Subtype/Link/A<</Type/Action/S/URI/URI(#1)>>}%
  \relax
 }%
 \providecommand\@@endlink[0]{\pdfendlink}%
}%
\providecommand \url  [0]{\begingroup\@sanitize \@url }%
\providecommand \@url [1]{\endgroup\@href {#1}{\urlprefix}}%
\providecommand \urlprefix [0]{URL }%
\providecommand \Eprint[0]{\href }%
\@ifxundefined \urlstyle {%
  \providecommand \doi [1]{doi:\discretionary{}{}{}#1}%
}{%
  \providecommand \doi [0]{doi:\discretionary{}{}{}\begingroup
  \urlstyle{rm}\Url }%
}%
\providecommand \doibase [0]{http://dx.doi.org/}%
\providecommand \Doi[1]{\href{\doibase#1}}%
\providecommand \selectlanguage [0]{\@gobble}%
\providecommand \bibinfo [0]{\@secondoftwo}%
\providecommand \bibfield [0]{\@secondoftwo}%
\providecommand \translation [1]{[#1]}%
\providecommand \BibitemOpen[0]{}%
\providecommand \bibitemStop [0]{}%
\providecommand \bibitemNoStop [0]{.\EOS\space}%
\providecommand \EOS [0]{\spacefactor3000\relax}%
\providecommand \BibitemShut [1]{\csname bibitem#1\endcsname}%
\bibitem{Pikovsky_et_al:Book}%
  \BibitemOpen
  \bibfield{author}{%
  \bibinfo {author} {\bibfnamefont{A.~S.}\ \bibnamefont{Pikovsky}}, \bibinfo
  {author} {\bibfnamefont{M.~G.}\ \bibnamefont{Rosenblum}},\ and\ \bibinfo
  {author} {\bibfnamefont{J.}~\bibnamefont{Kurths}},\ }%
  \emph{\bibinfo {title} {Synchronization: A Universal Concept in Nonlinear
  Sciences}}\ (\bibinfo {publisher} {Cambridge University Press},\ \bibinfo
  {address} {New York},\ \bibinfo {year} {2001})\BibitemShut{NoStop}%
\bibitem{Mosekilde_et_al:2002}%
  \BibitemOpen
  \bibfield{author}{%
  \bibinfo {author} {\bibfnamefont{E.}~\bibnamefont{Mosekilde}}, \bibinfo
  {author} {\bibfnamefont{D.}~\bibnamefont{Postnov}},\ and\ \bibinfo {author}
  {\bibfnamefont{Y.}~\bibnamefont{Maistrenko}},\ }%
  \emph{\bibinfo {title} {Chaotic Synchronization: Applications to Living
  Systems}}\ (\bibinfo {publisher} {World Scientific},\ \bibinfo {address}
  {Singapore},\ \bibinfo {year} {2002})\BibitemShut{NoStop}%
\bibitem{Strogatz:2003}%
  \BibitemOpen
  \bibfield{author}{%
  \bibinfo {author} {\bibfnamefont{S.~H.}\ \bibnamefont{Strogatz}},\ }%
  \emph{\bibinfo {title} {Sync: The Emerging Science of Spontaneous Order}}\
  (\bibinfo {publisher} {Hyperion Books},\ \bibinfo {address} {New York},\
  \bibinfo {year} {2003})\BibitemShut{NoStop}%
\bibitem{Balanov_et_al:2009}%
  \BibitemOpen
  \bibfield{author}{%
  \bibinfo {author} {\bibfnamefont{A.}~\bibnamefont{Balanov}}, \bibinfo
  {author} {\bibfnamefont{N.}~\bibnamefont{Janson}}, \bibinfo {author}
  {\bibfnamefont{D.}~\bibnamefont{Postnov}},\ and\ \bibinfo {author}
  {\bibfnamefont{O.}~\bibnamefont{Sosnovtseva}},\ }%
  \emph{\bibinfo {title} {Synchronization: From Simple to Complex}}\ (\bibinfo
  {publisher} {Springer},\ \bibinfo {address} {New York},\ \bibinfo {year}
  {2009})\BibitemShut{NoStop}%
\bibitem{Cumin_Unsworth:2007}%
  \BibitemOpen
  \bibfield{author}{%
  \bibinfo {author} {\bibfnamefont{D.}~\bibnamefont{Cumin}}\ and\ \bibinfo
  {author} {\bibfnamefont{C.~P.}\ \bibnamefont{Unsworth}},\ }%
  \bibfield{title}{%
  \enquote{\bibinfo {title} {Generalising the {K}uramoto model for the study of
  neuronal synchronization in the brain},}\ }%
  \bibfield{journal}{%
  \bibinfo {journal} {Physica D}\ }%
  \textbf{\bibinfo {volume} {226}},\ \bibinfo {pages} {181--196} (\bibinfo
  {month} {Feb.}\ \bibinfo {year} {2007})\BibitemShut{NoStop}%
\bibitem{Tass_book:1999}%
  \BibitemOpen
  \bibfield{author}{%
  \bibinfo {author} {\bibfnamefont{P.~A.}\ \bibnamefont{Tass}},\ }%
  \emph{\bibinfo {title} {Phase Resetting in Medicine and Biology. Stochastic
  Modelling and Data Analysis}}\ (\bibinfo {publisher} {Springer-Verlag},\
  \bibinfo {address} {Berlin},\ \bibinfo {year} {1999})\BibitemShut{NoStop}%
\bibitem{Milton-Jung-03}%
  \BibitemOpen
  \emph{\bibinfo {title} {Epilepsy as a Dynamic Disease}},\ edited by\ \bibinfo
  {editor} {\bibfnamefont{J.}~\bibnamefont{Milton}}\ and\ \bibinfo {editor}
  {\bibfnamefont{P.}~\bibnamefont{Jung}}\ (\bibinfo {publisher} {Springer},\
  \bibinfo {address} {Berlin},\ \bibinfo {year} {2003})\BibitemShut{NoStop}%
\bibitem{Buzhaki-Draguhn-04}%
  \BibitemOpen
  \bibfield{author}{%
  \bibinfo {author} {\bibfnamefont{G.}~\bibnamefont{Buzs\'aki}}\ and\ \bibinfo
  {author} {\bibfnamefont{A.}~\bibnamefont{Draguhn}},\ }%
  \bibfield{title}{%
  \enquote{\bibinfo {title} {Neuronal oscillations in cortical networks},}\ }%
  \bibfield{journal}{%
  \bibinfo {journal} {Science}\ }%
  \textbf{\bibinfo {volume} {304}},\ \bibinfo {pages} {1926--1929} (\bibinfo
  {year} {2004})\BibitemShut{NoStop}%
\bibitem{Batista_et_al:2010}%
  \BibitemOpen
  \bibfield{author}{%
  \bibinfo {author} {\bibfnamefont{C.A.S.}\ \bibnamefont{Batista}}, \bibinfo
  {author} {\bibfnamefont{S.R.}\ \bibnamefont{Lopes}}, \bibinfo {author}
  {\bibfnamefont{R.~L.}\ \bibnamefont{Viana}},\ and\ \bibinfo {author}
  {\bibfnamefont{A.~M.}\ \bibnamefont{Batista}},\ }%
  \bibfield{title}{%
  \enquote{\bibinfo {title} {Delayed feedback control of bursting
  synchronization in a scale-free neuronal network},}\ }%
  \bibfield{journal}{%
  \bibinfo {journal} {Neural Networks}\ }%
  \textbf{\bibinfo {volume} {23}},\ \bibinfo {pages} {114--124} (\bibinfo
  {month} {Jan.}\ \bibinfo {year} {2010})\BibitemShut{NoStop}%
\bibitem{Park_et_al:2011}%
  \BibitemOpen
  \bibfield{author}{%
  \bibinfo {author} {\bibfnamefont{C.}~\bibnamefont{Park}}, \bibinfo {author}
  {\bibfnamefont{R.~M.}\ \bibnamefont{Worth}},\ and\ \bibinfo {author}
  {\bibfnamefont{L.~L.}\ \bibnamefont{Rubchinsky}},\ }%
  \bibfield{title}{%
  \enquote{\bibinfo {title} {Neural dynamics in parkinsonian brain: The
  boundary between synchronized and nonsynchronized dynamics},}\ }%
  \bibfield{journal}{%
  \bibinfo {journal} {Phys. Rev. E}\ }%
  \textbf{\bibinfo {volume} {83}},\ \bibinfo {pages} {042901} (\bibinfo {month}
  {Oct.}\ \bibinfo {year} {2011})\BibitemShut{NoStop}%
\bibitem{Chkhenkeli-78}%
  \BibitemOpen
  \bibfield{author}{%
  \bibinfo {author} {\bibfnamefont{S.~A.}\ \bibnamefont{Chkhenkeli}},\ }%
  \bibfield{journal}{%
  \bibinfo {journal} {Bull. of Georgian Academy of Sciences}\ }%
  \textbf{\bibinfo {volume} {90}},\ \bibinfo {pages} {406--411} (\bibinfo
  {year} {1978})\BibitemShut{NoStop}%
\bibitem{Chkhenkeli-03}%
  \BibitemOpen
  \bibfield{author}{%
  \bibinfo {author} {\bibfnamefont{S.~A.}\ \bibnamefont{Chkhenkeli}},\ }%
  \enquote{\bibinfo {title} {Direct deep brain stimulation: {F}irst steps
  towards the feedback control of seizures},}\ in\ \emph{\bibinfo {booktitle}
  {Epilepsy as a Dynamic Disease}},\ \bibinfo {editor} {edited by\ \bibinfo
  {editor} {\bibfnamefont{J.}~\bibnamefont{Milton}}\ and\ \bibinfo {editor}
  {\bibfnamefont{P.}~\bibnamefont{Jung}}}\ (\bibinfo {publisher} {Springer},\
  \bibinfo {address} {Berlin},\ \bibinfo {year} {2003})\ pp.\ \bibinfo {pages}
  {249--261}\BibitemShut{NoStop}%
\bibitem{Benabid_et_al:1991}%
  \BibitemOpen
  \bibfield{author}{%
  \bibinfo {author} {\bibfnamefont{A.~L.}\ \bibnamefont{Benabid}}, \bibinfo
  {author} {\bibfnamefont{P.}~\bibnamefont{Pollak}}, \bibinfo {author}
  {\bibfnamefont{C.}~\bibnamefont{Gervason}}, \bibinfo {author}
  {\bibfnamefont{D.}~\bibnamefont{Hoffmann}}, \bibinfo {author}
  {\bibfnamefont{D~M.}\ \bibnamefont{Gao}}, \bibinfo {author}
  {\bibfnamefont{M.}~\bibnamefont{Hommel}}, \bibinfo {author}
  {\bibfnamefont{J.~E.}\ \bibnamefont{Perret}},\ and\ \bibinfo {author}
  {\bibfnamefont{J.}~\bibnamefont{de~Rougemont}},\ }%
  \bibfield{title}{%
  \enquote{\bibinfo {title} {Long-term suppression of tremor by chronic
  stimulation of the ventral intermediate thalamic nucleus},}\ }%
  \bibfield{journal}{%
  \bibinfo {journal} {Lancet.}\ }%
  \textbf{\bibinfo {volume} {337}},\ \bibinfo {pages} {403--406} (\bibinfo
  {month} {feb}\ \bibinfo {year} {1991})\BibitemShut{NoStop}%
\bibitem{Kringelbach-07}%
  \BibitemOpen
  \bibfield{author}{%
  \bibinfo {author} {\bibfnamefont{M.~L.}\ \bibnamefont{Kringelbach}}, \bibinfo
  {author} {\bibfnamefont{N.}~\bibnamefont{Jenkinson}}, \bibinfo {author}
  {\bibfnamefont{S.~L.~F.}\ \bibnamefont{Owen}},\ and\ \bibinfo {author}
  {\bibfnamefont{T.~Z.}\ \bibnamefont{Aziz}},\ }%
  \bibfield{title}{%
  \enquote{\bibinfo {title} {Translational principles of deep brain
  stimulation},}\ }%
  \bibfield{journal}{%
  \bibinfo {journal} {Nat Rev Neurosci}\ }%
  \textbf{\bibinfo {volume} {8}},\ \bibinfo {pages} {623--635} (\bibinfo {year}
  {2007})\BibitemShut{NoStop}%
\bibitem{Bronstein-11}%
  \BibitemOpen
  \bibfield{author}{%
  \bibinfo {author} {\bibfnamefont{J.~M.}\ \bibnamefont{Bronstein}}, \bibinfo
  {author} {\bibfnamefont{M.}~\bibnamefont{Tagliati}}, \bibinfo {author}
  {\bibfnamefont{R.L.}\ \bibnamefont{Alterman}}, \bibinfo {author}
  {\bibfnamefont{A.M.}\ \bibnamefont{Lozano}}, \bibinfo {author}
  {\bibfnamefont{J.}~\bibnamefont{Volkmann}}, \bibinfo {author}
  {\bibfnamefont{A.}~\bibnamefont{Stefani}}, \bibinfo {author}
  {\bibfnamefont{F.B.}\ \bibnamefont{Horak}}, \bibinfo {author}
  {\bibfnamefont{M.S.}\ \bibnamefont{Okun}}, \bibinfo {author}
  {\bibfnamefont{K.D.}\ \bibnamefont{Foote}}, \bibinfo {author}
  {\bibfnamefont{P.}~\bibnamefont{Krack}}, \bibinfo {author}
  {\bibfnamefont{R.}~\bibnamefont{Pahwa}}, \bibinfo {author}
  {\bibfnamefont{J.M.}\ \bibnamefont{Henderson}}, \bibinfo {author}
  {\bibfnamefont{M.I.}\ \bibnamefont{Hariz}}, \bibinfo {author}
  {\bibfnamefont{R.A.}\ \bibnamefont{Bakay}}, \bibinfo {author}
  {\bibfnamefont{A.}~\bibnamefont{Rezai}}, \bibinfo {author}
  {\bibfnamefont{W.I.}\ \bibnamefont{{Marks, Jr}}}, \bibinfo {author}
  {\bibfnamefont{E.}~\bibnamefont{Moro}}, \bibinfo {author}
  {\bibfnamefont{J.L.}\ \bibnamefont{Vitek}}, \bibinfo {author}
  {\bibfnamefont{F.M.}\ \bibnamefont{Weaver}}, \bibinfo {author}
  {\bibfnamefont{R.E.}\ \bibnamefont{Gross}},\ and\ \bibinfo {author}
  {\bibfnamefont{M.R.}\ \bibnamefont{DeLong}},\ }%
  \bibfield{title}{%
  \enquote{\bibinfo {title} {Deep brain stimulation for parkinson disease: an
  expert consensus and review of key issues},}\ }%
  \bibfield{journal}{%
  \bibinfo {journal} {Arch Neurol.}\ }%
  \textbf{\bibinfo {volume} {68}},\ \bibinfo {pages} {165} (\bibinfo {year}
  {2011})\BibitemShut{NoStop}%
\bibitem{Hammond_et_al:2008}%
  \BibitemOpen
  \bibfield{author}{%
  \bibinfo {author} {\bibfnamefont{C.}~\bibnamefont{Hammond}}, \bibinfo
  {author} {\bibfnamefont{R.}~\bibnamefont{Ammari}}, \bibinfo {author}
  {\bibfnamefont{B.}~\bibnamefont{Bioulac}},\ and\ \bibinfo {author}
  {\bibfnamefont{L.}~\bibnamefont{Garcia}},\ }%
  \bibfield{title}{%
  \enquote{\bibinfo {title} {Latest view on the mechanism of action of deep
  brain stimulation},}\ }%
  \bibfield{journal}{%
  \bibinfo {journal} {Mov. Disord.}\ }%
  \textbf{\bibinfo {volume} {23}},\ \bibinfo {pages} {2111--2121} (\bibinfo
  {month} {nov}\ \bibinfo {year} {2008})\BibitemShut{NoStop}%
\bibitem{Rosin_et_al:2011}%
  \BibitemOpen
  \bibfield{author}{%
  \bibinfo {author} {\bibfnamefont{B.}~\bibnamefont{Rosin}}, \bibinfo {author}
  {\bibfnamefont{M.}~\bibnamefont{Slovik}}, \bibinfo {author}
  {\bibfnamefont{R.}~\bibnamefont{Mitelman}}, \bibinfo {author}
  {\bibfnamefont{M.}~\bibnamefont{Rivlin-Etzion}}, \bibinfo {author}
  {\bibfnamefont{S.~N.}\ \bibnamefont{Haber}}, \bibinfo {author}
  {\bibfnamefont{Z.}~\bibnamefont{Israel}}, \bibinfo {author}
  {\bibfnamefont{E.}~\bibnamefont{Vaadia}},\ and\ \bibinfo {author}
  {\bibfnamefont{H.}~\bibnamefont{Bergman}},\ }%
  \bibfield{title}{%
  \enquote{\bibinfo {title} {Closed-loop deep brain stimulation is superior in
  ameliorating parkinsonism},}\ }%
  \bibfield{journal}{%
  \bibinfo {journal} {{ Neuron}}\ }%
  \textbf{\bibinfo {volume} {{ 72}}},\ \bibinfo {pages} {370--384} (\bibinfo
  {month} {Oct.}\ \bibinfo {year} {2011})\BibitemShut{NoStop}%
\bibitem{Bereny_et_al:2012}%
  \BibitemOpen
  \bibfield{author}{%
  \bibinfo {author} {\bibfnamefont{A.}~\bibnamefont{Ber\'enyi}}, \bibinfo
  {author} {\bibfnamefont{M.}~\bibnamefont{Belluscio}}, \bibinfo {author}
  {\bibfnamefont{D.}~\bibnamefont{Mao}},\ and\ \bibinfo {author}
  {\bibfnamefont{G.}~\bibnamefont{Buzs\'aki}},\ }%
  \bibfield{title}{%
  \enquote{\bibinfo {title} {Closed-loop control of epilepsy by transcranial
  electrical stimulation},}\ }%
  \bibfield{journal}{%
  \bibinfo {journal} {{ Science}}\ }%
  \textbf{\bibinfo {volume} {{ 337}}},\ \bibinfo {pages} {735--737} (\bibinfo
  {month} {Aug.}\ \bibinfo {year} {2012})\BibitemShut{NoStop}%
\bibitem{Tass:2003}%
  \BibitemOpen
  \bibfield{author}{%
  \bibinfo {author} {\bibfnamefont{P.~A.}\ \bibnamefont{Tass}},\ }%
  \bibfield{title}{%
  \enquote{\bibinfo {title} {A model of desynchronizing deep brain stimulation
  with a demand-controlled coordinated reset of neural subpopulations},}\ }%
  \bibfield{journal}{%
  \bibinfo {journal} {Biol. Cybern.}\ }%
  \textbf{\bibinfo {volume} {89}},\ \bibinfo {pages} {81--88} (\bibinfo {month}
  {Aug.}\ \bibinfo {year} {2003})\BibitemShut{NoStop}%
\bibitem{Lysyansky:2011}%
  \BibitemOpen
  \bibfield{author}{%
  \bibinfo {author} {\bibfnamefont{B.}~\bibnamefont{Lysyansky}}, \bibinfo
  {author} {\bibfnamefont{O.~V.}\ \bibnamefont{Popovych}},\ and\ \bibinfo
  {author} {\bibfnamefont{P.~A.}\ \bibnamefont{Tass}},\ }%
  \bibfield{title}{%
  \enquote{\bibinfo {title} {Desynchronizing anti-resonance effect of m : n
  on-off coordinated reset stimulation},}\ }%
  \bibfield{journal}{%
  \bibinfo {journal} {J. Neural Eng.}\ }%
  \textbf{\bibinfo {volume} {8}},\ \bibinfo {pages} {036019} (\bibinfo {month}
  {Jun.}\ \bibinfo {year} {2011})\BibitemShut{NoStop}%
\bibitem{Rosenblum_Pikovsky_a:2004}%
  \BibitemOpen
  \bibfield{author}{%
  \bibinfo {author} {\bibfnamefont{M.~G.}\ \bibnamefont{Rosenblum}}\ and\
  \bibinfo {author} {\bibfnamefont{A.~S.}\ \bibnamefont{Pikovsky}},\ }%
  \bibfield{title}{%
  \enquote{\bibinfo {title} {Controlling synchronization in an ensemble of
  globally coupled oscillators},}\ }%
  \bibfield{journal}{%
  \bibinfo {journal} {Phys. Rev. Lett.}\ }%
  \textbf{\bibinfo {volume} {92}},\ \bibinfo {pages} {114102} (\bibinfo {month}
  {Mar.}\ \bibinfo {year} {2004})\BibitemShut{NoStop}%
\bibitem{Rosenblum_Pikovsky_b:2004}%
  \BibitemOpen
  \bibfield{author}{%
  \bibinfo {author} {\bibfnamefont{M.~G.}\ \bibnamefont{Rosenblum}}\ and\
  \bibinfo {author} {\bibfnamefont{A.~S.}\ \bibnamefont{Pikovsky}},\ }%
  \bibfield{title}{%
  \enquote{\bibinfo {title} {Delayed feedback control of collective synchrony:
  an approach to suppression of pathological brain rhythms},}\ }%
  \bibfield{journal}{%
  \bibinfo {journal} {Phys. Rev. E}\ }%
  \textbf{\bibinfo {volume} {70}},\ \bibinfo {pages} {041904} (\bibinfo {year}
  {2004})\BibitemShut{NoStop}%
\bibitem{Popovych_et_al:2006}%
  \BibitemOpen
  \bibfield{author}{%
  \bibinfo {author} {\bibfnamefont{O.~V.}\ \bibnamefont{Popovych}}, \bibinfo
  {author} {\bibfnamefont{C.}~\bibnamefont{Hauptmann}},\ and\ \bibinfo {author}
  {\bibfnamefont{P.~A.}\ \bibnamefont{Tass}},\ }%
  \bibfield{title}{%
  \enquote{\bibinfo {title} {Control of neuronal synchrony by nonlinear delayed
  feedback},}\ }%
  \bibfield{journal}{%
  \bibinfo {journal} {Biol. Cybern.}\ }%
  \textbf{\bibinfo {volume} {95}},\ \bibinfo {pages} {69--85} (\bibinfo {month}
  {Jul.}\ \bibinfo {year} {2006})\BibitemShut{NoStop}%
\bibitem{Tukhlina_et_al:2007}%
  \BibitemOpen
  \bibfield{author}{%
  \bibinfo {author} {\bibfnamefont{N.}~\bibnamefont{Tukhlina}}, \bibinfo
  {author} {\bibfnamefont{M.~G.}\ \bibnamefont{Rosenblum}}, \bibinfo {author}
  {\bibfnamefont{A.~S.}\ \bibnamefont{Pikovsky}},\ and\ \bibinfo {author}
  {\bibfnamefont{J.}~\bibnamefont{Kurths}},\ }%
  \bibfield{title}{%
  \enquote{\bibinfo {title} {Feedback suppression of neural synchrony by
  vanishing stimulation},}\ }%
  \bibfield{journal}{%
  \bibinfo {journal} {Phys. Rev. E}\ }%
  \textbf{\bibinfo {volume} {75}},\ \bibinfo {pages} {011918} (\bibinfo {year}
  {2007})\BibitemShut{NoStop}%
\bibitem{Pyragas-Popovych-Tass-07}%
  \BibitemOpen
  \bibfield{author}{%
  \bibinfo {author} {\bibfnamefont{K.}~\bibnamefont{Pyragas}}, \bibinfo
  {author} {\bibfnamefont{O.~V.}\ \bibnamefont{Popovych}},\ and\ \bibinfo
  {author} {\bibfnamefont{P.~A.}\ \bibnamefont{Tass}},\ }%
  \bibfield{title}{%
  \enquote{\bibinfo {title} {Controlling synchrony in oscillatory networks with
  a separate stimulation-registration setup},}\ }%
  \bibfield{journal}{%
  \bibinfo {journal} {Europhys. Lett.}\ }%
  \textbf{\bibinfo {volume} {80}},\ \bibinfo {pages} {40002} (\bibinfo {year}
  {2007})\BibitemShut{NoStop}%
\bibitem{Popovych_et_al:2008}%
  \BibitemOpen
  \bibfield{author}{%
  \bibinfo {author} {\bibfnamefont{O.~V.}\ \bibnamefont{Popovych}}, \bibinfo
  {author} {\bibfnamefont{C.}~\bibnamefont{Hauptmann}},\ and\ \bibinfo {author}
  {\bibfnamefont{P.~A.}\ \bibnamefont{Tass}},\ }%
  \bibfield{title}{%
  \enquote{\bibinfo {title} {Impact of nonlinear delayed feedback on
  synchronized oscillators},}\ }%
  \bibfield{journal}{%
  \bibinfo {journal} {J. Biol. Phys.}\ }%
  \textbf{\bibinfo {volume} {34}},\ \bibinfo {pages} {367--379} (\bibinfo
  {month} {Nov.}\ \bibinfo {year} {2008})\BibitemShut{NoStop}%
\bibitem{Kobayashi_Kori:2009}%
  \BibitemOpen
  \bibfield{author}{%
  \bibinfo {author} {\bibfnamefont{Y.}~\bibnamefont{Kobayashi}}\ and\ \bibinfo
  {author} {\bibfnamefont{H.}~\bibnamefont{Kori}},\ }%
  \bibfield{title}{%
  \enquote{\bibinfo {title} {Design principle of multi-cluster and
  desynchronized states in oscillatory media via nonlinear global feedback},}\
  }%
  \bibfield{journal}{%
  \bibinfo {journal} {New J. Phys.}\ }%
  \textbf{\bibinfo {volume} {11}},\ \bibinfo {pages} {033018} (\bibinfo {month}
  {Mar.}\ \bibinfo {year} {2009})\BibitemShut{NoStop}%
\bibitem{Luo_et_al:2009}%
  \BibitemOpen
  \bibfield{author}{%
  \bibinfo {author} {\bibfnamefont{M.}~\bibnamefont{Luo}}, \bibinfo {author}
  {\bibfnamefont{Y.}~\bibnamefont{Wu}},\ and\ \bibinfo {author}
  {\bibfnamefont{J.}~\bibnamefont{Peng}},\ }%
  \bibfield{title}{%
  \enquote{\bibinfo {title} {Washout filter aided mean field feedback
  desynchronization in an ensemble of globally coupled neural oscillators},}\
  }%
  \bibfield{journal}{%
  \bibinfo {journal} {Biol. Cybern.}\ }%
  \textbf{\bibinfo {volume} {101}},\ \bibinfo {pages} {241--246} (\bibinfo
  {month} {Sep.}\ \bibinfo {year} {2009})\BibitemShut{NoStop}%
\bibitem{Popovych_Tass:2010}%
  \BibitemOpen
  \bibfield{author}{%
  \bibinfo {author} {\bibfnamefont{O.~V.}\ \bibnamefont{Popovych}}\ and\
  \bibinfo {author} {\bibfnamefont{P.~A.}\ \bibnamefont{Tass}},\ }%
  \bibfield{title}{%
  \enquote{\bibinfo {title} {Synchronization control of interacting oscillatory
  ensembles by mixed nonlinear delayed feedback},}\ }%
  \bibfield{journal}{%
  \bibinfo {journal} {Phys. Rev. E}\ }%
  \textbf{\bibinfo {volume} {82}},\ \bibinfo {pages} {026204} (\bibinfo {month}
  {Aug.}\ \bibinfo {year} {2010})\BibitemShut{NoStop}%
\bibitem{Luo_Xu:2011}%
  \BibitemOpen
  \bibfield{author}{%
  \bibinfo {author} {\bibfnamefont{M.}~\bibnamefont{Luo}}\ and\ \bibinfo
  {author} {\bibfnamefont{J.}~\bibnamefont{Xu}},\ }%
  \bibfield{title}{%
  \enquote{\bibinfo {title} {Suppression of collective synchronization in a
  system of neural groups with washout-filter-aided feedback},}\ }%
  \bibfield{journal}{%
  \bibinfo {journal} {Neural Networks}\ }%
  \textbf{\bibinfo {volume} {24}},\ \bibinfo {pages} {538--543} (\bibinfo
  {month} {Aug.}\ \bibinfo {year} {2011})\BibitemShut{NoStop}%
\bibitem{Franci_et_al:2011}%
  \BibitemOpen
  \bibfield{author}{%
  \bibinfo {author} {\bibfnamefont{A.}~\bibnamefont{Franci}}, \bibinfo {author}
  {\bibfnamefont{A.}~\bibnamefont{Chaillet}},\ and\ \bibinfo {author}
  {\bibfnamefont{W.}~\bibnamefont{Pasillas-L\'epine}},\ }%
  \bibfield{title}{%
  \enquote{\bibinfo {title} {Existence and robustness of phase-locking in
  coupled {K}uramoto oscillators under mean-field feedback},}\ }%
  \bibfield{journal}{%
  \bibinfo {journal} {Automatica}\ }%
  \textbf{\bibinfo {volume} {47}},\ \bibinfo {pages} {1193--1202} (\bibinfo
  {month} {Jun.}\ \bibinfo {year} {2011})\BibitemShut{NoStop}%
\bibitem{Franci_et_al:2012}%
  \BibitemOpen
  \bibfield{author}{%
  \bibinfo {author} {\bibfnamefont{A.}~\bibnamefont{Franci}}, \bibinfo {author}
  {\bibfnamefont{A.}~\bibnamefont{Chaillet}}, \bibinfo {author}
  {\bibfnamefont{E.}~\bibnamefont{Panteley}},\ and\ \bibinfo {author}
  {\bibfnamefont{F.}~\bibnamefont{Lamnabhi-Lagarrigue}},\ }%
  \bibfield{title}{%
  \enquote{\bibinfo {title} {Desynchronization and inhibition of {K}uramoto
  oscillators by scalar mean-field feedback},}\ }%
  \bibfield{journal}{%
  \bibinfo {journal} {Math. Control Signals Syst.}\ }%
  \textbf{\bibinfo {volume} {24}},\ \bibinfo {pages} {169--217} (\bibinfo
  {month} {Apr.}\ \bibinfo {year} {2012})\BibitemShut{NoStop}%
\bibitem{Hauptmann_et_al:2007}%
  \BibitemOpen
  \bibfield{author}{%
  \bibinfo {author} {\bibfnamefont{C.}~\bibnamefont{Hauptmann}}, \bibinfo
  {author} {\bibfnamefont{O.~V.}\ \bibnamefont{Popovych}},\ and\ \bibinfo
  {author} {\bibfnamefont{P.~A.}\ \bibnamefont{Tass}},\ }%
  \bibfield{title}{%
  \enquote{\bibinfo {title} {Demand-controlled desynchronization of oscillatory
  networks by means of a multisite delayed feedback stimulation},}\ }%
  \bibfield{journal}{%
  \bibinfo {journal} {Comp. Visual. Sci.}\ }%
  \textbf{\bibinfo {volume} {10}},\ \bibinfo {pages} {71--78} (\bibinfo {month}
  {Jun.}\ \bibinfo {year} {2007})\BibitemShut{NoStop}%
\bibitem{Omelchenko_et_al:2008}%
  \BibitemOpen
  \bibfield{author}{%
  \bibinfo {author} {\bibfnamefont{O.~E.}\ \bibnamefont{Omel`chenko}}, \bibinfo
  {author} {\bibfnamefont{C.}~\bibnamefont{Hauptmann}}, \bibinfo {author}
  {\bibfnamefont{Yu.~L.}\ \bibnamefont{Maistrenko}},\ and\ \bibinfo {author}
  {\bibfnamefont{P.~A.}\ \bibnamefont{Tass}},\ }%
  \bibfield{title}{%
  \enquote{\bibinfo {title} {Collective dynamics of globally coupled phase
  oscillators under multisite delayed feedback stimulation},}\ }%
  \bibfield{journal}{%
  \bibinfo {journal} {Physica D}\ }%
  \textbf{\bibinfo {volume} {237}},\ \bibinfo {pages} {365--384} (\bibinfo
  {month} {Mar.}\ \bibinfo {year} {2008})\BibitemShut{NoStop}%
\bibitem{Guo_Rubin:2011}%
  \BibitemOpen
  \bibfield{author}{%
  \bibinfo {author} {\bibfnamefont{Y.}~\bibnamefont{Guo}}\ and\ \bibinfo
  {author} {\bibfnamefont{J.~E.}\ \bibnamefont{Rubin}},\ }%
  \bibfield{title}{%
  \enquote{\bibinfo {title} {Multi-site stimulation of subthalamic nucleus
  diminishes thalamocortical relay errors in a biophysical network model},}\ }%
  \bibfield{journal}{%
  \bibinfo {journal} {Neural Networks}\ }%
  \textbf{\bibinfo {volume} {24}},\ \bibinfo {pages} {602--616} (\bibinfo
  {month} {Aug.}\ \bibinfo {year} {2011})\BibitemShut{NoStop}%
\bibitem{Rosenblum-Tukhlina-Pikovsky-Cimponeriu-06}%
  \BibitemOpen
  \bibfield{author}{%
  \bibinfo {author} {\bibfnamefont{M.~G.}\ \bibnamefont{Rosenblum}}, \bibinfo
  {author} {\bibfnamefont{N.}~\bibnamefont{Tukhlina}}, \bibinfo {author}
  {\bibfnamefont{A.~S.}\ \bibnamefont{Pikovsky}},\ and\ \bibinfo {author}
  {\bibfnamefont{L.}~\bibnamefont{Cimponeriu}},\ }%
  \bibfield{title}{%
  \enquote{\bibinfo {title} {Delayed feedback suppression of collective
  rhythmic activity in a neuronal ensemble},}\ }%
  \bibfield{journal}{%
  \bibinfo {journal} {Int. J. of Bifurcation and Chaos}\ }%
  \textbf{\bibinfo {volume} {16}},\ \bibinfo {pages} {1989--1999} (\bibinfo
  {year} {2006})\BibitemShut{NoStop}%
\bibitem{Tukhlina_Rosenblum:2008}%
  \BibitemOpen
  \bibfield{author}{%
  \bibinfo {author} {\bibfnamefont{N.}~\bibnamefont{Tukhlina}}\ and\ \bibinfo
  {author} {\bibfnamefont{M.~G.}\ \bibnamefont{Rosenblum}},\ }%
  \bibfield{title}{%
  \enquote{\bibinfo {title} {Feedback suppression of neural synchrony in two
  interacting populations by vanishing stimulation},}\ }%
  \bibfield{journal}{%
  \bibinfo {journal} {J. Biol. Phys.}\ }%
  \textbf{\bibinfo {volume} {34}},\ \bibinfo {pages} {301--314} (\bibinfo
  {month} {Jul.}\ \bibinfo {year} {2008})\BibitemShut{NoStop}%
\bibitem{Ott-Antonsen-08}%
  \BibitemOpen
  \bibfield{author}{%
  \bibinfo {author} {\bibfnamefont{E.}~\bibnamefont{Ott}}\ and\ \bibinfo
  {author} {\bibfnamefont{Th.~M.}\ \bibnamefont{Antonsen}},\ }%
  \bibfield{title}{%
  \enquote{\bibinfo {title} {Low dimensional behavior of large systems of
  globally coupled oscillators},}\ }%
  \bibfield{journal}{%
  \bibinfo {journal} {CHAOS}\ }%
  \textbf{\bibinfo {volume} {18}},\ \bibinfo {pages} {037113} (\bibinfo {year}
  {2008})\BibitemShut{NoStop}%
\bibitem{Ott-Antonsen-09}%
  \BibitemOpen
  \bibfield{author}{%
  \bibinfo {author} {\bibfnamefont{E.}~\bibnamefont{Ott}}\ and\ \bibinfo
  {author} {\bibfnamefont{Th.~M.}\ \bibnamefont{Antonsen}},\ }%
  \bibfield{title}{%
  \enquote{\bibinfo {title} {{Long time evolution of phase oscillator
  systems}},}\ }%
  \bibfield{journal}{%
  \bibinfo {journal} {{CHAOS}}\ }%
  \textbf{\bibinfo {volume} {{19}}},\ \bibinfo {pages} {{023117}} (\bibinfo
  {year} {{2009}})\BibitemShut{NoStop}%
\bibitem{PhysRevLett.97.213902}%
  \BibitemOpen
  \bibfield{author}{%
  \bibinfo {author} {\bibfnamefont{S.}~\bibnamefont{Schikora}}, \bibinfo
  {author} {\bibfnamefont{P.}~\bibnamefont{H\"ovel}}, \bibinfo {author}
  {\bibfnamefont{H.-J.}\ \bibnamefont{W\"unsche}}, \bibinfo {author}
  {\bibfnamefont{E.}~\bibnamefont{Sch\"oll}},\ and\ \bibinfo {author}
  {\bibfnamefont{F.}~\bibnamefont{Henneberger}},\ }%
  \bibfield{title}{%
  \enquote{\bibinfo {title} {All-optical noninvasive control of unstable steady
  states in a semiconductor laser},}\ }%
  \bibfield{journal}{%
  \Doi{10.1103/PhysRevLett.97.213902}{\bibinfo {journal} {Phys. Rev. Lett.}}\
  }%
  \textbf{\bibinfo {volume} {97}},\ \bibinfo {pages} {213902} (\bibinfo {month}
  {Nov}\ \bibinfo {year} {2006}),\
  \url{http://link.aps.org/doi/10.1103/PhysRevLett.97.213902}\BibitemShut{NoStop}%
\bibitem{Kaloust_Qu:1997}%
  \BibitemOpen
  \bibfield{author}{%
  \bibinfo {author} {\bibfnamefont{J.}~\bibnamefont{Kaloust}}\ and\ \bibinfo
  {author} {\bibfnamefont{Z.}~\bibnamefont{Qu}},\ }%
  \bibfield{title}{%
  \enquote{\bibinfo {title} {Robust control design for nonlinear uncertain
  systems with an unknown time-varying control direction},}\ }%
  \bibfield{journal}{%
  \bibinfo {journal} {IEEE Trans. Autom. Cntr.}\ }%
  \textbf{\bibinfo {volume} {42}},\ \bibinfo {pages} {393--399} (\bibinfo
  {month} {Mar.}\ \bibinfo {year} {1997})\BibitemShut{NoStop}%
\bibitem{Xu_Yan:2004}%
  \BibitemOpen
  \bibfield{author}{%
  \bibinfo {author} {\bibfnamefont{J.X.}\ \bibnamefont{Xu}}\ and\ \bibinfo
  {author} {\bibfnamefont{R.}~\bibnamefont{Yan}},\ }%
  \bibfield{title}{%
  \enquote{\bibinfo {title} {Iterative learning control design without a priori
  knowledge of the control direction},}\ }%
  \bibfield{journal}{%
  \bibinfo {journal} {Automatica}\ }%
  \textbf{\bibinfo {volume} {40}},\ \bibinfo {pages} {1803--1809} (\bibinfo
  {month} {Oct.}\ \bibinfo {year} {2004})\BibitemShut{NoStop}%
\bibitem{Kaloust_Qu:1995}%
  \BibitemOpen
  \bibfield{author}{%
  \bibinfo {author} {\bibfnamefont{J.}~\bibnamefont{Kaloust}}\ and\ \bibinfo
  {author} {\bibfnamefont{Z.}~\bibnamefont{Qu}},\ }%
  \bibfield{title}{%
  \enquote{\bibinfo {title} {Continuous robust control design for nonlinear
  uncertain systems without a priori knowledge of control direction},}\ }%
  \bibfield{journal}{%
  \bibinfo {journal} {IEEE Trans. Autom. Cntr.}\ }%
  \textbf{\bibinfo {volume} {40}},\ \bibinfo {pages} {276--282} (\bibinfo
  {month} {Feb.}\ \bibinfo {year} {1995})\BibitemShut{NoStop}%
\bibitem{Ge_et_al:2008}%
  \BibitemOpen
  \bibfield{author}{%
  \bibinfo {author} {\bibfnamefont{S.~S.}\ \bibnamefont{Ge}}, \bibinfo {author}
  {\bibfnamefont{C.}~\bibnamefont{Yang}},\ and\ \bibinfo {author}
  {\bibfnamefont{T.~H.}\ \bibnamefont{Lee}},\ }%
  \bibfield{title}{%
  \enquote{\bibinfo {title} {Adaptive robust control of a class of nonlinear
  strict-feedback discrete-time systems with unknown control directions},}\ }%
  \bibfield{journal}{%
  \bibinfo {journal} {Syst. Control Lett.}\ }%
  \textbf{\bibinfo {volume} {57}},\ \bibinfo {pages} {888--895} (\bibinfo
  {month} {Nov.}\ \bibinfo {year} {2008})\BibitemShut{NoStop}%
\bibitem{Liu_Huang:2008}%
  \BibitemOpen
  \bibfield{author}{%
  \bibinfo {author} {\bibfnamefont{L.}~\bibnamefont{Liu}}\ and\ \bibinfo
  {author} {\bibfnamefont{J.}~\bibnamefont{Huang}},\ }%
  \bibfield{title}{%
  \enquote{\bibinfo {title} {Global robust output regulation of lower
  triangular systems with unknown control direction},}\ }%
  \bibfield{journal}{%
  \bibinfo {journal} {Automatica}\ }%
  \textbf{\bibinfo {volume} {44}},\ \bibinfo {pages} {1278--â1284} (\bibinfo
  {month} {May}\ \bibinfo {year} {2008})\BibitemShut{NoStop}%
\bibitem{Wen_Ren:2011}%
  \BibitemOpen
  \bibfield{author}{%
  \bibinfo {author} {\bibfnamefont{Y.}~\bibnamefont{Wen}}\ and\ \bibinfo
  {author} {\bibfnamefont{X.}~\bibnamefont{Ren}},\ }%
  \bibfield{title}{%
  \enquote{\bibinfo {title} {Neural networks-based adaptive control for
  nonlinear time-varying delays systems with unknown control direction},}\ }%
  \bibfield{journal}{%
  \bibinfo {journal} {IEEE Trans. Neural Networks}\ }%
  \textbf{\bibinfo {volume} {22}},\ \bibinfo {pages} {1599--1612} (\bibinfo
  {month} {Oct.}\ \bibinfo {year} {2011})\BibitemShut{NoStop}%
\bibitem{Bartolini_et_al:2009}%
  \BibitemOpen
  \bibfield{author}{%
  \bibinfo {author} {\bibfnamefont{G.}~\bibnamefont{Bartolini}}, \bibinfo
  {author} {\bibfnamefont{A.}~\bibnamefont{Pisano}},\ and\ \bibinfo {author}
  {\bibfnamefont{E.}~\bibnamefont{Usai}},\ }%
  \bibfield{title}{%
  \enquote{\bibinfo {title} {On the second-order sliding mode control of
  nonlinear systems with uncertain control direction},}\ }%
  \bibfield{journal}{%
  \bibinfo {journal} {Automatica}\ }%
  \textbf{\bibinfo {volume} {45}},\ \bibinfo {pages} {2982--2985} (\bibinfo
  {month} {Dec.}\ \bibinfo {year} {2009})\BibitemShut{NoStop}%
\bibitem{Note1}%
  \BibitemOpen
  \bibinfo {note} {A second order system in the strict feedback form is
  represented as $\protect \mathaccentV {dot}05Fx_1=f_1(x_1)+g_1x_2$, $\protect
  \mathaccentV {dot}05Fx_2=f_2(x_1,x_2) + g_2 u$, where $g_{1,2}$ are unknown
  parameters. A $(2+m)^{th}$ order system in the normal form is represented as
  $\protect \mathaccentV {dot}05Fx_1=x_2$, $\protect \mathaccentV
  {dot}05Fx_2=f(x_1,x_2,y) +g u$, $\protect \mathaccentV
  {dot}05Fy=q(x_1,x_2,y)$, where $y\in \protect \mathbb {R}^{m}$ is the vector
  of additional states (which describe internal dynamics of the feedback loop;
  see Ref.~\cite {Khalil:2002} for more details).}\BibitemShut{Stop}%
\bibitem{Oliveira_et_al:2010}%
  \BibitemOpen
  \bibfield{author}{%
  \bibinfo {author} {\bibfnamefont{T.~R.}\ \bibnamefont{Oliveira}}, \bibinfo
  {author} {\bibfnamefont{A.~J.}\ \bibnamefont{Peixoto}},\ and\ \bibinfo
  {author} {\bibfnamefont{L.}~\bibnamefont{Hsu}},\ }%
  \bibfield{title}{%
  \enquote{\bibinfo {title} {Sliding mode control of uncertain multivariable
  nonlinear systems with unknown control direction via switching and monitoring
  function},}\ }%
  \bibfield{journal}{%
  \bibinfo {journal} {IEEE Trans. Autom. Cntr.}\ }%
  \textbf{\bibinfo {volume} {55}},\ \bibinfo {pages} {1028--1034} (\bibinfo
  {year} {2010})\BibitemShut{NoStop}%
\bibitem{Pikovsky_Ruffo:1999}%
  \BibitemOpen
  \bibfield{author}{%
  \bibinfo {author} {\bibfnamefont{A.}~\bibnamefont{Pikovsky}}\ and\ \bibinfo
  {author} {\bibfnamefont{S.}~\bibnamefont{Ruffo}},\ }%
  \bibfield{title}{%
  \enquote{\bibinfo {title} {Finite-size effects in a population of interacting
  oscillators},}\ }%
  \bibfield{journal}{%
  \bibinfo {journal} {Phys. Rev. E.}\ }%
  \textbf{\bibinfo {volume} {59}},\ \bibinfo {pages} {1633} (\bibinfo {year}
  {1999})\BibitemShut{NoStop}%
\bibitem{Khalil:2002}%
  \BibitemOpen
  \bibfield{author}{%
  \bibinfo {author} {\bibfnamefont{H.~K.}\ \bibnamefont{Khalil}},\ }%
  \emph{\bibinfo {title} {Nonlinear Control Systems}}\ (\bibinfo {publisher}
  {Prentic Hall},\ \bibinfo {address} {New Jersey},\ \bibinfo {year}
  {2002})\BibitemShut{NoStop}%
\end{thebibliography}

%

\end{document}